\documentclass[a4paper]{amsart}
\usepackage{amscd}
\def\opn#1 {\operatorname{#1}}
\def\dopn#1 {
 \def\mYname{\operatorname{#1}}\expandafter\let\csname#1\endcsname=\mYname}

\def\up#1{\def\arg{#1} \sp{\hbox{$\arg$}}}
\def\dn#1{\def\arg{#1} \sb{\hbox{$\arg$}}}
\def\br#1{
 \ifx#1<\gdef\Br##1>{\left<##1\right>}\else
 \ifx#1(\gdef\Br##1){\left(##1\right)}\else
 \ifx#1[\gdef\Br##1]{\left[##1\right]}\else
 \ifx#1\{\gdef\Br##1\}{\left\{##1\right\}}\else
 \ifx#1|\gdef\Br##1|{\left|##1\right|}\else
 \ifx#1\|\gdef\Br##1\|{\left\|##1\right\|}\else
 \errmessage{MYMAC: Bad bracket!}\fi\fi\fi\fi\fi\fi
 \Br}

\let\too=\xrightarrow
\def\seq{\subseteq}
\def\cj{\overline}
\def\ul{\underline}

\newtheorem{thm}{Theorem}[subsection]
\newtheorem{cor}[thm]{Corollary}

\newtheorem{prp}[thm]{Proposition}
          
\theoremstyle{definition}
\newtheorem{dfn}[thm]{Definition}
\theoremstyle{remark}
       
\newtheorem{rem}{Remark}    
\newtheorem{rmn}[thm]{Remark} 
\newtheorem{rems}{Remarks}    
\newtheorem{rmns}[thm]{Remarks} 

\long\def\CAR#1#2\NIL{#1}
\long\def\Brm#1\Erm{
 \edef\nxt{\CAR#1\relax\NIL}
 \expandafter\ifx\nxt(
 \begin{rems} #1 \end{rems}\else
 \begin{rem} #1 \end{rem}\fi
}
\long\def\Brmn#1 #2\Erm{
 \edef\nxt{\CAR#2\relax\NIL}
 \expandafter\ifx\nxt(
 \begin{rmns}\label{#1} #2 \end{rmns}\else
 \begin{rmn}\label{#1} #2 \end{rmn}\fi
}

\numberwithin{equation}{subsection}
\def\nt{\cr} 
\def\Beq#1\Eeq{\begin{equation*} #1 \end{equation*}}
\def\Beqn#1 #2\Eeq{\begin{equation}#2 \label{#1} \end{equation}}
\def\Bml#1\Eml{\begin{multline*} #1 \end{multline*}}
\def\Bmln#1 #2\Eml{\begin{multline}#2 \label{#1} \end{multline}}
\def\Bal#1\Eal{\begin{align*} #1 \end{align*}}
\def\Baln#1 #2\Eal{\begin{align}\label{#1} #2 \end{align}}
\def\Bgt#1\Egt{\begin{gather*} #1 \end{gather*}}
\def\Bgtn#1 #2\Egt{\begin{gather}\label{#1} #2 \end{gather}}
\def\Bea#1\Eea{\begin{eqnarray*} #1 \end{eqnarray*}}
\def\Bean#1 #2\Eea{\begin{eqnarray} #2 \label{#1}\end{eqnarray}}
\def\Bcd#1\Ecd{\[\begin{CD} #1 \end{CD}\]}
\def\Bcdn#1 #2\Ecd{
 \begin{equation}\begin{CD}#2 \label{#1}\end{CD}\end{equation}}
\def\bysame{\leavevmode\hbox to3em{\hrulefill}\,} 

\begin{document}
\def\even{{\mathbf0}}
\def\odd{{\mathbf1}}
\def\seven{_\even}
\def\sodd{_\odd}
\def\sevR{_{\even,\mathbb R}}
\def\sodR{_{\odd,\mathbb R}}
\def\O{\mathop{\mathcal O}\nolimits}
\def\Oinf{\O_\infty}
\def\M{\mathop{\mathcal M}\nolimits}
\dopn L
\def\Cau{^{\opn Cau }}
\def\sol{^{\opn sol }}
\def\rdmm{\mathbb R^{d+1}}
\newdimen\mYd  \newbox\mYbox
\def\dcj#1{\def\mYarg{#1}
 \setbox\mYbox=\hbox{$\mYarg$}\mYd=\wd\mYbox
 \vbox{
  \offinterlineskip
  \hbox{\vrule width\mYd height1pt}
  \vskip1pt
  \box\mYbox
}}
\def\P#1;{{\mathcal P}#1;}
\def\Space{\opn space }
\dopn Spec
\dopn dim
\dopn An
\dopn CS
\dopn Re
\dopn Im
\dopn i
\dopn t
\dopn S
\dopn T
\dopn pr
\def\Ort{{\opn O }}
\def\cV{{\mathcal V}}
\def\new{_{\mathbf s}}
\def\CMPref#1{\cite[#1]{[CMP1]}}
\def\Higgs{_{\opn H }}
\def\Spi{_{\opn S }}
\def\Ferm{_{\opn F }}
\def\Gauge{_{\opn G }}
\def\doublearrow{\vcenter{\offinterlineskip\hbox{$\to$}\hbox{$\to$}}}
\def\ztwo{$\mathbb Z_2$}
\def\ztN{\mathbb Z_2^N}
\def\bot{\mathop{\textstyle\bigotimes}\nolimits}
\hyphenation{equi-va-len-ce}

\def\BerSmf{\text{\bf Ber-Smfs}}
\def\Smfs{\text{\bf Smfs}}
\def\Gr{\text{\bf Gr}}
\def\Sets{\text{\bf Sets}}
\def\CatC{\text{\bf C}} 
\def\Mfs{\text{\bf Mfs}} 

\title[Supergeometry and QFT]
{
 Supergeometry and Quantum Field Theory, or:
\linebreak
 What is a Classical Configuration?
}

\author{T. Schmitt}
\thanks{
 Special thanks to the late German Democratic Republic who
 made this research possible by continuous financial support
 over ten years
}
\begin{abstract}
We discuss of the conceptual difficulties connected with the
anticommutativity of classical fermion fields, and we argue
that the "space" of all classical configurations of a
model with such fields should be described as an infinite-dimensional
supermanifold $M$.
We discuss the two main approaches to supermanifolds,
and we examine the reasons why many physicists tend to prefer the Rogers
approach although the Berezin-Kostant-Leites approach is the more
fundamental one. We develop the infinite-dimensional variant of the latter,
and we show that the superfunctionals considered in \cite{[CMP1]}
are nothing but superfunctions on $M$.
We propose a programme for future mathematical work, which applies to any
classical field model with fermion fields. A part of this programme
will be implemented in the successor paper \cite{[CMP3]}.
\end{abstract}
\maketitle

\tableofcontents

\section*{Introduction}

Although this paper logically continues \cite{[CMP1]}, it can be
read independently.

We begin with a discussion of the conceptual difficulties connected with the
anticommutativity of classical (=unquantized) fermion fields, and we argue
that only a supergeometric approach allows a convincing description of
these fields; indeed, the "space" of all classical configurations of a
model with such fields should be described as an infinite-dimensional
supermanifold.

Before explicating this, we discuss the two main approaches to supermanifolds,
starting with the approach of Berezin, Leites, Kostant et al, together with
the hermitian modification proposed in \cite{[HERM]}.
We then review the second, alternative approach proposed by B. deWitt and
A. Rogers and followed by quite a few authors
and its connection to the first one as well as the "interpolating"
approach of Molotkov (\cite{[Mol 84]}),
and we examine the reasons why many physicists tend to prefer it
although the Berezin-Kostant-Leites approach is the more fundamental one.

In particular, we show that the deWitt-Rogers supermanifolds suffer from the
same conceptual shortcomings as the naive configuration notion in classical
field theory with anticommuting fields.

The deWitt-Rogers approach is intimately connected with an
"interim solution" of the conceptual difficulties mentioned: one considers
"$B$-valued configurations", i. e. instead of the usual real or complex
numbers, one uses a Grassmann algebra $B$ (or some relative of it) as target
for the classical fields, providing in this way the apparently needed
"anticommuting values". However, this approach is suspicious for the very same
reasons as the deWitt-Rogers supermanifold approach; in particular, in view
of the arbitrariness in the choice of $B$. Therefore we go one step beyond and
re-interpret $B$-valued configurations as {\em families of configurations}
parametrized by supermanifolds. Once this is done, it is natural to state the
question for a universal family and a corresponding moduli space -- an idea
which directly leads to the infinite-dimensional configuration supermanifold
mentioned.

Hence, we have to develop the infinite-dimensional variant
of the Berezin-Kostant-Leites approach, basing on previous work of
the author. In fact, we will show that the superfunctionals
considered in \cite{[CMP1]} are nothing but {\em superfunctions on an
infinite-dimensional supermanifold}. If the model is purely bosonic then a
superfunctional is an ordinary functional, i. e. a function the domain of
definition of which is a function space.

Generalizing the philosophy of the paper \cite{[Ber/Mar]} from quantum
mechanics onto quantum field theory, we propose a programme for future
mathematical work, which applies to any classical field model with
fermion fields, and which aims at
understanding its mathematical structure. A part of this programme
will be implemented in the successor paper \cite{[CMP3]}.


\section{Supergeometry and its relationship to quantum field theory}

\subsection{Geometric models of quantum field theory
and classical configurations}

The modern realistic quantum field theoretic models, like
quantum electrodynamics, Salam-Weinberg electroweak theory,
quantum chromodynamics, and all these unified in the standard model,
have achieved remarkable success in predicting experimental results,
in one case even with an accuracy up to 11 digits. Even for a
mathematician, which turns up his nose about the mathematical and
logical status ot these models, this should be strong
evidence that these models have something to do with nature, and
that they should be taken seriously.
Even if it should turn out that they do not produce
rigorous quantum theories in Wightman's sense, they have proven to give
at least a kind of semiclassical limit plus quite a few quantum corrections of
a more fundamental theory (provided the latter exists), and therefore
it remains a promising task to clear up the mathematical structure, and
to put it into a consistent framework as far as possible.
Moreover, there exist phenomenological models in hadron and
nuclear physics which are rather unlikely to be implementable as
rigorous quantum theories, due to non-renormalizability; nevertheless, they
give information, and thus should be taken serious.

The models mentioned start with formulating a {\em classical} field theory
of fields on Minkowski space $\mathbb R^4$, and the hypothetical final theory
is thought to arise by quantization of
the classical theory in a similar sense as quantum mechanics arises by
quantization of point mechanics. Of course, the concept of quantization
is problematic even in the latter situation; but Kostant-Souriau
geometric quantization gives at least one mathematically well-defined
procedure.

In the field theoretic situation, where we have infinitely many degrees
of freedom, geometric quantization in its usual form does no longer apply,
and there arises a gap between mathematics and physics:
The models which have been constructed rigorously (like the $P(\Phi)_d$ models,
the Thirring model, the Gross-Neveu model (cf. \cite{[Rivasseau]},
\cite{[Iagolnitzer]} and references therein)),
are only toy models in small space-time dimensions, while the realistic models
mentioned above are treated up to now only heuristically (this statement
is not affected
by the fact that some special features have been successfully treated by
rigourous mathematical methods, like e. g. the instanton solutions of
euclidian pure Yang-Mills theory).

A popular device to tackle the field quantization problem heuristically
is Feynman's path integral. Inspite of the fact that only some aspects
are mathematically understood (cf. e. g. Osterwalder-Schrader axiomatics
a la Glimm-Jaffe \cite{[Gl/Ja]}), and that only the bosonic part can be
understood as integration over a measure space, it has been very
successful in deriving computation rules by formal manipulation like
variable changing, integration by parts, standard Gaussian integrals.

Now the path integral is supposed to live on the space $M$ of classical
configurations $\Xi$ of the model: under some restrictions onto the
Lagrangian density ${\mathcal L}(\Xi)$, "each one" of them enters formally
with the weight $\exp(\i\int d^4x\;{\mathcal L}[\Xi](x)/\hbar)$ (or, in the
euclidian variant, minus instead of the imaginary unit).

Now, if fermion fields enter the game then a complication arises (and this
applies in the standard models both to the matter fields and to the
ghost fields in the Faddeev-Popov scheme): Any classical fermion field
$\Psi_\alpha$ has to be treated as anticommuting,
\Beqn AntiComm
 \br[\Psi_\alpha(x),\Psi_\beta(y)]_+=0
\Eeq
($x,y\in\mathbb R^4$ are space-time points; physicists use to say that
$\Psi_\alpha(x)$ is an "a-number"), and this makes the concept
of a "classical configuration" for fermion fields problematic.
("Fermion" refers always to the statistics, not to the spin.
The difference matters for ghost fields, which do not obey the Pauli Theorem.)

In \cite{[Ber2ndQ]}, Berezin developed a recipe to compute Gaussian functional
integrals over fermions in a heuristic way (actually, his recipe is
mathematically precise but applies only to rather "tame" Gaussian kernels;
however, the practical application is heuristic because the kernels are
far from being "tame"). Thus, in the standard models, one can often
circumvent the problem mentioned above by "throwing out" the fermions by
Gaussian integration, so that only bosonic configurations remain.
(However, the Faddeev-Popov scheme for the quantization of gauge theories
does the opposite: after some formal manipulations with the bosonic
functional integral, a determinant factor appears which is represented by
a fermionic Gaussian integral. The rest is Feynman perturbation theory.)

On the other hand, the problem of understanding the notion of
a classical configuration becomes urgent in supersymmetric models,
because -- at least in component field formulation --
supersymmetry simply does not work with the naive notion of
configuration with commuting fields, and the computations of physicists
would not make sense. Only anticommuting fields behave formally
correct.

Also, in the mathematical analysis of the BRST approach to the quantization
of constrained systems (cf. e. g. \cite{[Kostant-Sternberg 87]}), the
fermionic, anticommuting nature of the ghost degrees of freedom
(ghost fields in the context of Yang-Mills theory) has to be taken into
account.

Contrary to naive expectations, the law \eqref{AntiComm} is not automatically
implemented by considering superfield models on supermanifolds:
The components $f_\mu(x)$ of a superfunction
\Beq
 f(x,\xi)=\sum f_\mu(x)\xi^\mu
\Eeq
are still $\mathbb R$- (or $\mathbb C$-) valued and therefore commuting
quantities while e. g. for the chiral superfield (cf. e. g.
\cite[Ch. V]{[We/Ba]}, therein called "scalar superfield")
\Beqn ChirSupF
 \Phi(y,\theta)=A(y)+2\theta\psi(y)+\theta\theta F(y),
\Eeq
the Weyl spinor $\psi(x)$, which describes a left-handed fermion,
has to satisfy the law \eqref{AntiComm} -- at least if non-linear expressions
in $\psi$ have to be considered.

Thus, in the framework of superfield models, it still remains
as obscure as in component field models how to implement \eqref{AntiComm}
mathematically. Nevertheless, we will see that
it is just supergeometry which is the key to an appropriate solution
of this problem. But before establishing the link, we review
the history and the main approaches to supergeometry.

Unfortunately, this problem has been mostly neglected in the existing
literature both on super differential geometry and on mathematical
modelling of field theories with fermions. In the worst case,
fermions were modelled by ordinary sections in spinor bundles, like
e.~g. in \cite{[DelSmo]}. (Of course, this is okay as long as all
things one does are
linear in the fermion fields. For instance, this applies to the
vast literature on the Dirac equation in a gauge field background.
On the other hand, if e.~g. the first Yang-Mills equation with
spinor current source or supersymmetry transformations have to
be considered then this approach becomes certainly inconsistent).


\subsection{Remarks on the history of supergeometry}

"Supermathematics", and, in particular, super differential geometry,
has its roots in the fact (which is familiar and unchallenged
among physicists, although not seldom ignored by mathematicians)
that in the framework of modern quantum field theory, which uses
geometrically formulated field theories, fermions are described
on a classical level by anticommuting fields. Indeed, practical
experience in heuristic computations showed that if keeping track
of the signs one can handle anticommuting fields formally "in
the same way" as commuting (bosonic) ones. The most striking
example for this is the Feynman path integral, or its
relative, the functional integral, for fermion fields;
cf. \cite{[Ber2ndQ]}. This experience led F. A. Berezin to the conclusion
that "there exists a non-trivial analogue of analysis in which
anticommuting variables appear on equal footing with the usual
variables" (quotation from \cite{[Ber2]}), and he became
the pioneer in constructing this theory. The book \cite{[Ber2]}
which summarizes Berezin's work in this field, reflects supergeometry
"in statu nascendi".

The papers \cite{[Kos]}, \cite{[Lei1]}, and the book \cite{[Lei2]}
give systematic, methodologically closed presentations of the
foundations of supergemetry; cf. \cite{[Lei1]} also for more
information on the history of the subject.

In the meantime, supergeometry turned to more special questions,
in particular, to the investigation of super-analogues of important
classical structures. As examples from the vast literature, let us
mention Serre and Mekhbout duality on complex supermanifolds
(\cite{[Pen]}), Lie superalgebras
and their representations (see e. g. \cite{[Kac]}, \cite{[Scheu]}),
integration theory on supermanifolds (\cite{[Bern/Lei1]},
\cite{[Bern/Lei2]}, \cite{[Rot]}), deformations of complex supermanifolds
(\cite{[Vain]}), integrability of CR structures (\cite{[Sch2]}, \cite{[Sch3]}),
the investigation of twistor geometry and Penrose transform in the super
context (\cite{[Man1]} and references therein). (Of course,
the quotations are by no means complete.) Manin's book \cite{[Man1]} seems
to be one of the most far-reaching efforts to apply supergeometry to
classical field models in a mathematical rigorous way.

Also, one should mention the "Manin programme" \cite{[Man2]} which
calls for treating the "odd dimensions" of super on equal footing
with the ordinary "even dimensions" as well as with the "arithmetic"
or "discrete dimensions" of number theory, and for considering
all three types of dimensions in their dialectic relations to
each other.

Now if speaking on the history of supergeometry one should also
mention the alternative approach of B. deWitt and A. Rogers to supermanifolds,
which was followed (and often technically modified) by several
authors (cf. e.g. \cite{[Ja/Pi]}, \cite{[Vl/Vo]}, \cite{[DeW2]}),
and the theory of Molotkov \cite{[Mol 84]} which in some sense unifies
both approaches in an elegant way and allows also infinite-dimensional
supermanifolds. We will comment this below.

\subsection{A short account of the Berezin-Leites-Kostant approach}

Roughly speaking, a {\em finite-dimensional supermanifold} ({\em smf})
$X$ is a "geometrical object" on which there exist locally, together
with the usual even coordinates $x_1,\dots,x_m$, odd,
anticommuting coordinates $\xi_1,\dots,\xi_n$.

In order to implement this mathematically, one defines a {\em
(smooth) superdomain} $U$ of dimension $m|n$ as a ringed space
$U = (\Space(U),\Oinf)$ where $\Space(U)\seq\mathbb R^m$ is an open
subset, and the structure sheaf is given by
\Beqn FDStrSheaf
 \Oinf(\cdot):=
 C^\infty(\cdot)\otimes_{\mathbb R} \mathbb R[\xi_1,\dots,\xi_n]
\Eeq
where $\xi_1,\dots,\xi_n$ is a sequence of Grassmann variables. Thus,
$\Oinf(V)$ for open $V\seq U$ is the algebra of all formal sums
\Beqn SupFctExp
 f(x|\xi)=\sum
  f_{\mu_1\cdots\mu_n}(x)\xi^{\mu_1}\cdots\xi^{\mu_n}
 =\sum f_\mu(x)\xi^\mu
\Eeq
where the sum runs over all $2^n$ tuples
$\mu=(\mu_1,\dots,\mu_n)\in\mathbb Z_2^n$,
and the $f_\mu(x)$ are smooth functions $V\to\mathbb R$. One interprets
\eqref{SupFctExp} as the Taylor expansion of $f(x|\xi)$ w. r. to the
Grassmann variables; due to their anticommmutativity, it terminates after
$2^n$ terms.

Here and in the following, we use the subscript $\infty$ at the structure
sheaf in order to emphasize that we are considering the smooth variant
of the theory; the symbol $\O$ without subscript is reserved to complex-
and real-analytic structure sheaves, as already considered in \CMPref{3.5}.
Note that in the finite-dimensional situation, the definitions of the
complex- and real-analytic variants of \eqref{FDStrSheaf}
are straightforward. The "globality theorems" like
\cite[Prps. 2.4.1, 2.4.2, etc.]{[Kos]}, or the quirk of \cite{[Kos]} of
defining morphisms by homomorphisms of the algebra of global superfunctions,
do no longer work in these situations. However, if one instead follows the
standard framework of ringed spaces there is no genuine obstackle to a
satisfactory calculus.

Although the use of a {\em real} Grassmann algebra in
\eqref{FDStrSheaf} looks quite natural, it is in fact highly problematic;
we will discuss this in \ref{HermStuff} below.

Now one defines a {\em supermanifold (smf)} as a ringed
space $X=(\Space(X),\allowbreak\Oinf)$ with the underlying space paracompact
and Hausdorff which is locally isomorphic to a superdomain. (Kostants
definition is easily seen to be equivalent to this one.)
Morphisms of smfs are just morphisms of ringed spaces.
We will not mention here the various differential-geometric ramifications;
in section \ref{InfDSmfs}, we will develop the infinite-dimensional variant
of some of them. We only note that every smf determines an
{\em underlying $C^\infty$ manifold}, and that, conversely, every
$C^\infty$ manifold can be viewed as smf.

For later use we note also that to every
finite-dimensional \ztwo-graded vector space $V=V\seven\oplus V\sodd$
there belongs a {\em linear superspace} (often called also "affine superspace",
in analogy with the habits of algebraic geometry)
$\L(V)$ which has the even part $V\seven$ as underlying space and
$C^\infty(\cdot)\otimes \Lambda V\sodd^*$ as structure sheaf. Here
$\Lambda V\sodd^*$ is the exterior algebra over the dual of the odd part;
due to the dualization, one has a natural embedding of the whole dual into
the global superfunctions,
\Beqn VInOV
 V^*\seq\Oinf(\L(V)),
\Eeq
and $V^*$ is just the space of {\em linear superfunctions}. This
procedure is just a superization of the fairly trivial fact that
every finite-dimensional vector space can be viewed as smooth manifold.

It is of outstanding importance that a morphism from an arbitrary
smf $X$ to a linear smf $\L(V)$ is known by knowing the coordinate pullbacks:
Fix a basis $u_1,\dots,u_k\in V^*\seven$, \ \ $\xi_1,\dots,\xi_l\in V^*\sodd$
of the dual $V^*$; because of \eqref{VInOV}, these elements become
superfunctions $u_1,\dots,u_k,\xi_1,\dots,\xi_l\in \Oinf(\L(V))$.
It is reasonable to call this tuple a {\em coordinate system} on
$\L(V)$, and we have (cf. \cite[Thm. 2.1.7]{[Lei1]},
\cite[Thm. 3.1]{[SchReport]}, \cite[Cor. 3.3.2]{[HERM]}):

\begin{thm}\label{FinDimCoords}
Given superfunctions
$u'_1,\dots,u'_k\in\Oinf(X)\seven$, \ \
$\xi'_1,\dots,\xi'_l\in\Oinf(X)\sodd$
there exists a unique morphism $\mu: X\to\L(V)$ such that
$\mu^*(u_i)= u'_i$, \ \ $\mu^*(\xi_j)= \xi'_j$ for all $i,j$.
\qed\end{thm}

This Theorem makes the morphisms between the local models controllable.
One can give it a more invariant form:

\begin{cor}\label{FinDimPbChar}
Given an smf $X$, a finite-dimensional \ztwo-graded vector space $V$,
and an element $\hat\mu\in(\Oinf(X)\otimes V)\seven$
there exists a unique morphism $\mu: X\to\L(V)$ such that
for any $v\in V^*$ we have $\br<v,\hat\mu> = \mu^*(v)$ within
$\Oinf(X)$.
\qed\end{cor}

In this form, it will also generalize to
the infinite-dimensional situation (cf. Thm. \ref{CoordChar} below);
however, the real-analytic analogon $\O(\cdot)\otimes V$ of the sheaf
$\Oinf(\cdot)\otimes V$
will have to be replaced by a
bigger sheaf $\O^V(\cdot)$ of "$V$-valued superfunctions".

\Brm
Several authors, like e. g. \cite{[Mol 84]}, consider generalized frameworks
based on a $\ztN$ grading, so that the parity rule
as well as the first sign rule have to be applied to each degree $|\cdot|_j$
separately. Thus, a $\ztN$-graded algebra
$R = \bigoplus_{(i_1,\dots,i_N)\in\ztN} R_{(i_1,\dots,i_N)}$ is
{\em $\ztN$-commutative} iff
\Beq
 ba = (-1)^{\br|a|_1\br|b|_1+ \dots \br|a|_N\br|b|_N}ab
\Eeq
for homogeneous $a,b\in R$.
Some people claim that the use of such a structure is appropriate in order
to separate physical fermion field d.~o.~f.'s from ghost field d.~o.~f.'s.

However, as observed e.~g. in \cite[Ex. 6.15]{[Hen/Tei]},
one can work without such a generalization, thus saving at least some
notations: One may construct from $R$ a \ztwo-graded algebra $R\new$ by taking
the total degree,
\[
 (R\new)_{\mathbf k} := \bigoplus_{(i_1,\dots,i_N)\in\ztN:\
  i_1+\dots+i_N\equiv {\mathbf k}\mod 2} R_{(i_1,\dots,i_N)}
\]
for $\mathbf k=\mathbf 0,\mathbf 1$; thus $|a|\new = \sum_i |a|_i$.
Also, one modifies the multiplication law by
\[
 a\cdot\new b := (-1)\up{\sum_{i>j} \br|a|_i\br|b|_j}ab
\]
for homogeneous elements.
This new product is again associative;
if $R$ was $\ztN$-commutative then $R\new$ is
\ztwo-commutative. Analogous redefinitions can be made for modules and other
algebraic structures, so that there is no real necessity to consider such
generalized gradings.

In particular, instead of constructing the exterior differential on an
smf as even operator (cf. e.~g. \cite{[Kos]}), which leads to a
$\mathbb Z_2\times\mathbb Z_+$-graded
commutative algebra of differential forms, it is more appropriate to
use an odd exterior differential. This leads to an algebra of differential
forms which is still $\mathbb Z_2\times\mathbb Z_+$-graded but where only the
\ztwo degree produces sign factors. Such an approach does not only
simplify computations but it opens the way for a natural supergeometric
interpretation of exterior, interior, and Lie derivatives as vector fields
on a bigger smf, and it leads straightforwardly to the Bernstein-Leites
pseudodifferential forms. Cf. \cite{[Bern/Lei1]}, \cite{[SchReport]},
\cite{[IS2]}.
\Erm

\subsection{Supergeometry and physicists}

Supergeometry has both its sources and its justification mainly
in the classical field models used in quantum field theory. Nowadays
it is most used for the formulation and analysis of supersymmetric
field models in terms of superfields which live on a supermanifold
(cf. e. g. \cite{[We/Ba]}, \cite{[Man1]}). However, most of the
physicists who work in this field either do not make use of the
mathematical theory at all, or they use only fragments of it
(mainly the calculus of differential forms and connections, and
the Berezin integral over volume forms).

The first and most obvious reason for this is that one can do
formulation and perturbative analysis of superfield theories
without knowing anything of ringed spaces (one has to know how
to handle superfunctions $f(x^\mu,\theta^\alpha,\cj\theta^{\dot\alpha})$
which depend on commuting coordinates $x^\mu$ and anticommuting
coordinates $\theta^\alpha$, $\cj\theta^{\dot\alpha}$;
for most purposes it is inessential how this is mathematically
implemented).

The second reason is fairly obvious, too: Despite the invasion
of manifolds, fibre bundles, cohomology, Chern classes etc. into
mathematical physics -- the work with e. g. morphisms (instead
of "genuine maps" of sets) or group objects (instead of "genuine
groups") is not easy even for the mathematically well-educated
physicist. However, the difficulties are of purely psychological
nature; the algebraic geometer has to deal with quite similar
(and sometimes still far more abstract) structures since Grothendieck's
revolution in this area (one should mention here that also $C^\infty$
supergeometry has learned a lot from algebraic geometry).

But there are also two further, less obvious reasons. One of
them is a certain conceptual shortcoming of Berezin's theory
concerning the treatment of complex conjugation, which made it
almost inapplicable to physical models. In \cite{[HERM]}, the
present author made a proposal to modify this theory in order
to adapt it to the needs of quantum physics; cf. \ref{HermStuff} below for a
short account.

Finally, the fourth, and perhaps most severe
reason is the problem of modelling classical fermion fields mentioned
above. In particular, the Berezin-Kostant-Leites approach does not
provide anticommuting constants, which seem to be indispensable
(but are, as we will see in \ref{FamPhilo}, in fact not).

Although supergeometry is usually considered to be merely a tool
for the formulation of supersymmetric models in terms of superfields on
supermanifolds, its conceptual importance is much wider: while
the $x_i$ describe usual, "bosonic" degrees
of freedom the $\xi_j$ describe new, "fermionic" degrees of freedom
which are not encountered in classical mechanics. In fact,
{\em supergeometry is the adequate tool to describe the classical
limit of fermionic degrees of freedom}.

This point of view is pursued also in the paper \cite{[Ber/Mar]} of
Berezin and Marinov. However, there seems to be a minor conceptual
misunderstanding in that paper which we want to comment on:
The authors argue that one can describe the algebra of classical
observables of a spinning electron in an external field as a
Grassmann algebra $R=\mathbb C\br[\xi_1,\xi_2,\xi_3]$ (or, more precisely,
as the even, real part of $R$). Time evolution of observables
$f\in R$ is governed by $df/dt=\{H,f\}$ where $\{\cdot,\cdot\}$ is a
Poisson bracket making $R$ a Lie superalgebra, and $H\in R\seven$
is the Hamiltonian. In supergeometric language, the phase space
is the $0|3$-dimensional hermitian linear supermanifold (cf. \ref{HermStuff})
$\L(\mathbb R^{0|3})$ equipped with the symplectic structure
$\omega=(d\xi_1)^2+(d\xi_2)^2+(d\xi_3)^2$
which induces the Poisson bracket in $R=\Oinf(\L(\mathbb R^{0|3}))$.

Now the notion of "phase space trajectory" introduced in
\cite[2.1]{[Ber/Mar]} as a smooth map $\mathbb R\to R\sodd$, $t\mapsto\xi(t)$,
leads to confusion. Indeed, $t\mapsto\xi(t)$ describes the
time evolution of the observable $\xi(0)$ rather than that of
a state (that is, we have a "Heisenberg picture" at the unquantized
level).

In accordance with this, for the "phase space distribution"
$\rho$ of \cite[2.2]{[Ber/Mar]}, there is no sensible
notion of a $\delta$-distribution which would indicate a "pure
state". Thus, it should be stressed that in this approach -- as
well as in our one to be developed below -- there do not exist
"individual configurations" on the unquantized level.

With this point of view on \cite{[Ber/Mar]}, one may say that the
programme presented in section \ref{Program} below is the logical extension
of the philosophy of \cite{[Ber/Mar]} from classical mechanics onto
classical field theory.

For a systematic (but rather formal) treatment of supermanifolds as
configuration spaces of mechanical systems, cf. the book \cite{[Hen/Tei]}
and references therein.

In \ref{Lattice}, we will see how supermanifolds emerge naturally
as configuration spaces of lattice theories; the advantage of those is that
the finite-dimensional calculus is sufficient.

Now if we consider a {\em field} (instead of, say, a mass point)
then we will have infinitely many degrees of freedom. Roughly
speaking, the bosonic and fermionic field strengthes
$\Phi_i(x)$, $\Psi_j(x)$ for all space-time points $x$ are just the
coordinates of the configuration space.

The appearance of anticommuting functions on the
configuration space indicates that the latter should be understood
as an {\em infinite-dimensional supermanifold}. We will elaborate
this philosophy from \ref{FamPhilo} on.

However, the use of anticommuting variables is not restricted to the
unquantized theory: "Generating functionals" of states and operators in  Fock
space were invented by F. A. Berezin \cite{[Ber2ndQ]} thirty years
ago.  While in the purely bosonic case they are analytic
functions  on  the  one-particle  state  space $H$, their geometric
interpretation remained up to now obscure on the fermionic side:  they  are
just  elements of an "infinite-dimensional Grassmann  algebra" the
origin of which was not clear.

Infinite-dimensional supergeometry provides a satisfactory solution
of the riddle:  the Grassmann algebra mentioned
above  is  nothing  but  the algebra  of  superfunctions  on  the
$0|\infty$-dimensional supermanifold  $\L(H)$. More generally,
supergeometry allows to unify  bosonic and fermionic case, instead
of  considering  them separately, as Berezin did. Cf. \cite{[Rapallo]}
for a sketch; a detailed presentation will appear elsewhere.

\def\euc{_{\opn euc }}
\dopn J
\label{Schwinger}
Another natural appearance of supergeometry is in
quantized euclidian theory, which is given by its {\em Schwinger functions},
i. e. the vacuum expectation values of the quantized fields:
\Bal
 &S^{I|J}(X|Y) = S^{i_1,\dots,i_k|j_1,\dots,j_l}(x_1,\dots,x_k|y_1,\dots,y_l)
 :=
 \left<\hat\Phi_{i_1}(x_1)\cdots\hat\Phi_{i_k}(x_k)
 \hat\Psi_{j_1}(y_1)\cdots\hat\Psi_{j_l}(y_l)\right>_{\opn vac }
 \\
 &\quad= \frac 1N \int [D\Phi][D\Psi]
 \Phi_{i_1}(x_1)\cdots\Phi_{i_k}(x_k)\cdot \Psi_{j_1}(y_1)\cdots
 \Psi_{j_l}(y_l)
 \exp(\frac{-1}\hbar S\euc[\Phi|\Psi]).
\Eal
Here $S\euc[\Phi|\Psi]$ is the euclidian action which depends on
bosonic fields $\Phi_i$ and fermionic ones $\Psi_j$ (for notational
simplicity, we use real field components).

Usually, one assumes the Schwinger functions to be tempered distributions
defined on the whole space, $S^{I|J}\in{\mathcal S}'(\mathbb R^{d(k+l)})$,
and satisfying the Osterwalder-Schrader axioms
(cf. \cite[9.5.B]{[BigBog]}).
The Schwinger functions now are the coefficient functions
(cf. \CMPref{2.3}) of the {\em Euclidian generating functional}
\Bal
 Z\euc[\J^\Phi|\J^\Psi]
 &= \sum_{k,l\ge0}\frac 1{k!l!} \sum_{i_1,\dots,i_k,j_1,\dots,j_l}
 \int dXdY S^{I|J}(X|Y)\J^\Phi_{i_1}(x_1)\cdots\J^\Phi_{i_k}(x_k)
 \J^\Psi_{j_1}(y_1)\cdots\J^\Psi_{j_l}(y_l)
 \\
 &= \frac 1N \int [D\Phi][D\Psi]
 \exp\Bigl(\frac {-1}\hbar S\euc[\Phi|\Psi] + \int dx(\sum_i
\J^\Phi_i(x)\Phi_i(x)
  + \sum_j \J^\Psi_j(x)\Psi_j(x))\Bigr)
\Eal
where the functional variables $\J^\Phi,\J^\Psi$ are
(formal) "external sources"; note that $\J^\Phi$, $\J^\Psi$ are commuting
and anticommuting, respectively.

In the purely bosonic case,
$Z\euc[\J^\Phi]: {\mathcal S}(\mathbb R^d)\otimes V^*\to\mathbb C$ is the
characteristic
function of a measure on the euclidian configuration space
${\mathcal S}'(\mathbb R^d)\otimes V$ of the theory; here $V$ is the field
target
space as in \CMPref{2.2}. However, if fermionic fields are present then,
due to the antisymmetry of the Schwinger functions in the fermionic sector,
$Z\euc[\J^\Phi|\J^\Psi]$ cannot be interpreted as a map any longer.
Instead of this, it is at least a  formal power series in the sense of
\CMPref{2.3}, and thanks to the usual regularity condition on the Schwinger
functions, it becomes a superfunction
$Z\euc[\J^\Phi|\J^\Psi]\in\O(\L({\mathcal S}(\mathbb R^d)\otimes V^*))$.
(It is a challenging task to construct some superanalogon of measures, so that
$Z\euc[\J^\Phi|\J^\Psi]$ becomes the characteristic superfunction of a
supermeasure on the euclidian configuration supermanifold
$\L({\mathcal S}'(\mathbb R^d)\otimes V)$ of the theory.)

Apart from the last remark, analogous remarks apply also to the
(time-ordered) Green functions in Minkowski theory. Note, however, that,
although their existence is generally assumed, it does not follow from
the Wightman axioms; cf. also \cite[13.1.A]{[BigBog]}.

\subsection{Supergeometry and hermitian conjugation}\label{HermStuff}

Here, we give a short account of \cite{[HERM]}.

Usually, one uses the sign rule in the following form:

{\bf Sign Rule.}  Whenever in a multilinear expression standing
on the r.h.s.  of an equation two adjacent terms  $A,B$  are
interchanged  (with  respect to their position on the r. h. s.)  a  sign
$(-1)^{\br|A|\br|B|}$  occurs.

This rule has is origin in the commutation rule of Grassmann algebra,
and it is well-known whereever these are applied, e. g.
in supergeometry as well as in homological algebra,
differential geometry (in treating exterior forms), and algebraic topology.

However, the law of operator conjugation in the quantized theory,
$(AB)^* = B^*A^*$ yields in the classical limit the rule
$\cj{fg} = \cj g\cj f$ for superfunctions $f,g$ on a supermanifold.
This seems to contradict the sign rule, and the best way out of the
trouble is the following:

One applies the sign rule only to {\em complex} multilinear expressions.
In fact, one avoids the use of merely real-linear terms; that is, all
vector spaces the elements of which appear in multilinear terms should be
complex. With this first step, the contradiction is resolved, but there
remains incertainty. Thus, we have to do more.

Skew-linearity should appear only in the form of explicit hermitian
conjugation, and the latter is treated by:

{\bf Second Sign Rule.} If conjugation is applied to a bilinear
expression containing the adjacent terms $a,\ b$ (i. e. if conjugation
is resolved into termwise conjugation), either $a,\ b$ have to be rearranged
backwards, or the expression acquires the sign factor $(-1)^{\br|a|\br|b|}$.
Multilinear terms have to be treated iteratively.

These rules have consequences for the very definition of supermanifolds:
In a natural way, one defines a {\em hermitian ringed space} as a pair
$X = (\Space(X),\Oinf)$ where $\Space(X)$ is a topological space,
the structure sheaf $\Oinf$ is now a sheaf of {\it complex} \ztwo-graded
algebras which are equipped with a {\em hermitian conjugation}, i. e.
an antilinear involution $^-:\Oinf\to\Oinf$ which satisfies
\Beq
 \cj{fg}=\cj g\cj f \qquad\text{for $f,g\in\Oinf$}.
\Eeq
Now one defines a {\em hermitian superdomain} $U$ of dimension $m|n$ as
a hermitian ringed space $U = (\Space(U),\Oinf)$ where
$\Space(U)\seq\mathbb R^n$ is open again but
\Beq
 \Oinf(\cdot)=C_{\mathbb C}^\infty(\cdot)\otimes_{\mathbb C}\mathbb
C[\xi_1,\dots,\xi_n]
\Eeq
where $\xi_1,\dots,\xi_n$ is again a sequence of Grassmann variables. Thus,
$\Oinf(V)$ for open $V\seq U$ is the algebra of all formal sums
\eqref{SupFctExp} where, however, the smooth functions $f_\mu(x)$ are
now complex-valued. The hermitian conjugation is defined by the properties
\Beq
 \cj f = f \quad\text{for $f\in C_{\mathbb R}^\infty$}, \qquad
 \cj\xi_i=\xi_i \quad\text{for all $i$.}
\Eeq
Thus
\Beqn HermConjLaw
 \cj{f(x|\xi)} = \sum (-1)^{\br|\mu|(\br|\mu|-1)/2}\cj{f_\mu(x)}\xi^\mu.
\Eeq
Now one defines a {\em hermitian supermanifold} as a hermitian ringed
space with the underlying space paracompact
and Hausdorff which is locally isomorphic to a hermitian superdomain.
Note that the definition of the linear superspace belonging to a
finite-dimensional \ztwo-graded vector space $V=V\seven\oplus V\sodd$ also
changes:
\Beq
 \L(V) =(V\seven,\  C^\infty_{\mathbb C}(\cdot)
 \otimes_{\mathbb C} \Lambda V_{\mathbf1,\mathbb C}^*).
\Eeq
Thm. \ref{FinDimCoords} slightly changes because the coordinate pullbacks
now have to be required real:
$u'_1,\dots,u'_k\allowbreak \in\Oinf(X)\sevR$, \ \
$\xi'_1,\dots,\xi'_l\in\Oinf(X)_{\odd,\mathbb R}$.
Also, in Cor. \ref{FinDimPbChar}, the element $\hat\mu$ has to be required
real: $\hat\mu\in(\Oinf(X)\otimes V)\sevR$.

Cf. \cite{[HERM]} for the rest of the story.

\Brm
It is interesting to note that among the people who worked on
supercalculus questions, DeWitt was one of the few who
did not walk into the trap of "mathematical simplicity" concerning
real structures: Although in his book \cite{[DeW2]}  he does not formulate
the second sign rule, he actually works in the
hermitian framework instead of the traditional one from the
beginning. In particular, his "algebra of supernumbers"
$\Lambda_\infty$ is a hermitian algebra (cf. \CMPref{2.1} or \cite{[HERM]}),
and the "supervector space" introduced in \cite[1.4]{[DeW2]} is the
same as a free hermitian module over $\Lambda_\infty$.

Also, the book \cite{[Hen/Tei]}, being concerned with both classical
and quantum aspects of ghosts and physical fermions,
uses in fact a hermitian approach.
\Erm


\subsection{Sketch of the deWitt-Rogers approach}\label{Sketch}
The difficulties mentioned motivated physicists to look for
an alternative approach to the mathematical implementation of
supermanifolds. This approach was pioneered by B. deWitt \cite{[DeW2]} and
A. Rogers \cite{[Rog]}, and it was followed (and often technically modified)
by several authors (cf. e.g. \cite{[Ja/Pi]}, \cite{[Vl/Vo]}).
Cf. the book \cite{[Bruzzo]} for more on the history.

The basic idea consists in realizing $m|n$-dimensional
superspace as the topological space $B\seven^m\times B\sodd^n$ where
$B=B\seven\oplus B\sodd$ is a suitable topological \ztwo-commutative
algebra. Suitability is defined differently in each version of the theory,
but for the present discussion, the differences do not matter much. Here we
follow the original paper of Rogers \cite{[Rog]}, that is,
$B=\Lambda[\beta_1,\dots\beta_k]$ is a finite-dimensional Grassmann algebra.
Equipping $B$ with the norm $\br\|\sum c_\mu\beta^\mu\| := \sum \br|c_\mu|$,
it becomes a Banach algebra.

(We note that, from the point of view of the Berezin approach, it looks
suspicious that odd things appear as geometrical points. One should really
work better with the purely even object $B\seven^m\times(\Pi B\sodd)^n$ where
$\Pi$ is the parity shift symbol. In this form, $m|n$-dimensional superspace
will naturally appear also in  \ref{Ber2Rog} below.)

Now a map $f: U\to B$ where $U\seq B\seven^m\times B\sodd^n$ is open
is called {\em superdifferentiable} iff there exist maps
$D_if: U\to B$ such that
\Beq
 \lim \frac{\br\|f(x+y) - f(x) - \sum y_iD_if(x)\|}{\br\|y\|} =0
\Eeq
for $x=(x_1,\dots,x_{m+n})\in U$ and $y\to 0$ in $B\seven^m\times B\sodd^n$.

The problem with this definition is that,
while the even partial derivatives $D_1f,\dots,D_kf$ are uniquely determined,
the odd ones, $D_{m+1}f,\dots,D_{m+n}f$ are determined only up to a summand
of the form $g(x)\beta_1\cdots\beta_k$ where the map $g: U\to\mathbb R$ is
arbitrary (!). This is the first appearance of the "truncation
effects" which are typical for finite-dimensional $B$. They can be avoided
by using infinite-dimensional Grassmann algebras, but only at the price
of other technical and conceptual complications; so we stick here to
$k<\infty$.

Let $G^\infty(U)$ be the algebra of all infinitely often superdifferentiable
functions on $U$. More exactly, one defines inductively
$G^l(U)$ by letting $G^0(U)$ be just the set of all continuous maps
$U\to B$, and $G^{l+1}(U)$ be the set of all $f:U\to B$ which
are superdifferentiable such that all $D_if$ can be choosen in
$G^l(U)$; finally, $G^\infty(U):=\bigcap_{l>0} G^l(U)$.

In an obvious way, $G^\infty(U)$ is a \ztwo-graded commutative
algebra over $B$.

Note that since $B$ is a $2^k$-dimensional vector space, $f$ encodes
$2^k$ real-valued functions on $\mathbb R^N$ with $N:=2^{k-1}(m+n)$; and
superdifferentiability requires not only that these functions are
differentiable but that their derivatives satisfy a certain linear system
of relations, somewhat analogous to the Cauchy-Riemann relations in function
theory. The expansion \eqref{RogExpans} below is a consequence of
these relations.

The most obvious example for a superdifferentiable function is the projection
$x_i: B\seven^m\times B\sodd^n\to B$ from the $i$-th factor;
$x_i\in G^\infty(U)$ will play the r\^ole of the $i$-th coordinate.
Sometimes we will have to treat the even and the odd ones separately;
so we will also write
$(u_1,\dots,u_m,\xi_1,\dots,\xi_n):=(x_1,\dots,x_{m+n})$.

Now let $\epsilon: B\to\mathbb R$ be the unique algebra homomorphism; thus, for
every $a\in B$, its "soul" $a-\epsilon(a)$ is nilpotent, and $a$ is invertible
iff its "body", the number $\epsilon(a)$, is non-zero.

\Brm
In the various variants of the theory, one has always
such a unique body projection; however, if $B$ is an infinite-dimensional
algebra, the soul may be only topologically nilpotent.
\Erm

We get a "body projection" from to superspace to ordinary space:
\Beq
 \epsilon: B\seven^m\times B\sodd^n\to\mathbb R^m,\quad
 (u_1,\dots,u_m,\xi_1,\dots,\xi_n)\mapsto
 \br({\epsilon(u_1),\dots,\epsilon(u_m)}).
\Eeq
Let $U\seq B\seven^m\times B\sodd^n$ be open. Given
$f\in C^\infty(\epsilon(U))\otimes B$, i. e. a $B$-valued smooth function on
$\epsilon(U)$, we get by "Grassmann analytic continuation" an element
$z(f)\in G^\infty(U)$,
\Beq
 z(f)(u):= \sum_{\nu_1,\dots,\nu_m\ge0}
  \partial_1^{\nu_1}\cdots \partial_m^{\nu_m}f(\epsilon(u))
  \frac{(u_1-\epsilon(u_1))^{\nu_1}\cdots(u_m-\epsilon(u_m))^{\nu_1}}
   {\nu_1!\cdots\nu_m!};
\Eeq
the sum is actually finite. One gets an injective algebra homomorphism
$z: C^\infty(\epsilon(U))\to G^\infty(U)$, and it turns out that
$G^\infty(U)$ is generated as algebra by the union of the image of this
homomorphism with the set of coordinates $\{x_1,\dots,x_{m+n}\}$.
Indeed, one finds that every $f\in G^\infty(U)$ admits an expansion
\Beqn RogExpans
 f(u|\xi)=\sum
  z(f_{\mu_1\dots\mu_n})(u)\xi_1^{\mu_1}\cdots\xi_n^{\mu_n}
 =\sum z(f_\mu)(u)\xi^\mu
\Eeq
which not accidentally looks like \eqref{SupFctExp}, but here
with suitable $f_{\mu_1\dots\mu_n}\in C^\infty(\epsilon(U))$.
Unfortunately, here the truncation effects make their second appearance:
the $f_{\mu_1\dots\mu_n}$ are uniquely determined only for
$\mu_1+\cdots+\mu_n \le k$; otherwise, they are completely undetermined
since $\xi_1^{\mu_1}\cdots\xi_n^{\mu_n}$ vanishes anyway.

Thus, we get an epimorphism of \ztwo-graded algebras
\Beqn TheEpi
 C^\infty(\epsilon(U))\otimes B\otimes
 \Lambda[\xi_1,\dots,\xi_m]\to G^\infty(U),\quad
 f\otimes \xi_1^{\mu_1}\cdots\xi_n^{\mu_n}\mapsto
 z(f_{\mu_1\dots\mu_n})\xi_1^{\mu_1}\cdots\xi_n^{\mu_n}.
\Eeq
Its kernel is generated by all products
$\xi_1^{\mu_1}\cdots\xi_n^{\mu_n}$ with $\mu_1+\dots+\mu_n>k$; in
particular, \eqref{TheEpi} is an isomorphism iff $n\le k$.

A map $B\seven^m\times B\sodd^n\supseteq U\to B\seven^{m'}\times B\sodd^{n'}$,
where $U$ is open in the natural topology, is called $G^\infty$ iff its
$m'+n'$ component maps are $G^\infty$. Now, globalizing in a standard way,
one gets the notion of a {\em $G^\infty$ supermanifold} which is a
paracompact Hausdorff space together with an atlas of
$B\seven^m\times B\sodd^n$-valued charts such that the transition maps are
$G^\infty$.

On such a $G^\infty$ smf, there lives the sheaf $G^\infty(\cdot)$ of
$B$-valued functions of class $G^\infty$; one could also define
a $G^\infty$ smf equivalently as a paracompact Hausdorff ringed space
which is locally isomorphic to the model space
$(B\seven^m\times B\sodd^n,G^\infty(\cdot))$.

Now there exists a second reasonable notion of superfunctions
and supermanifolds: Let $H^\infty(U)$ be the subalgebra of all
$f\in G^\infty(U)$ for which the $f_{\mu_1\dots\mu_n}$ can be choosen
$\mathbb R$-valued. This is still a \ztwo-graded commutative algebra over
$\mathbb R$ but not over $B$, i. e. there are no longer anticommuting
constants.
Now \eqref{TheEpi} restricts to an algebra epimorphism
\Beqn TheRestEpi
 C^\infty(\epsilon(U))\otimes
 \Lambda[\xi_1,\dots,\xi_n]\to H^\infty(U).
\Eeq
Using $H^\infty$ transition maps instead of $G^\infty$ ones one gets the
notion of a {\em $H^\infty$ smf}.
On a $H^\infty$ smf, there lives the sheaf $H^\infty(\cdot)$ which consists
of $B$-valued functions. However, it is no longer a sheaf of
algebras over $B$ because $H^\infty(U)\cap B=\mathbb R$ within
$G^\infty(U)$ for any open $U$.

One could also define a $H^\infty$ smf equivalently as a paracompact
Hausdorff ringed space which is locally isomorphic to the model space
$(B\seven^m\times B\sodd^n,H^\infty(\cdot))$.

Now every $H^\infty$ supermanifold is also a $G^\infty$ supermanifold in
a natural way but not conversely; for the sheaves one finds
$G^\infty(\cdot) \cong H^\infty(\cdot)\otimes B$.

A further ramification arises if one uses instead of the natural topology on
$B\seven^m\times B\sodd^n$ the {\em deWitt topology} which is the inverse
image topology arising from $\epsilon: B\seven^m\times B\sodd^n\to\mathbb R^m$;
note that it is not Hausdorff. However, although it is much coarser than the
natural topology, the loss of information is smaller than one would think:
It follows from the expansion \eqref{RogExpans} that once $U\seq V$ are
open in the natural topology with $\epsilon(U)=\epsilon(V)$ the
restriction maps $G^\infty(V)\to G^\infty(U)$ and $H^\infty(V)\to H^\infty(U)$
are bijective.

So one defines a {\em $H^\infty$-deWitt supermanifold} as a paracompact space
together with an atlas of
$B\seven^m\times B\sodd^n$-valued charts, which are now local homeomorphisms
w. r. to the deWitt topology, and the transition maps are $H^\infty$.
Indeed, we will see that the smf arising from a Berezin smf is naturally of
this type.

\subsection{From Berezin to deWitt-Rogers}\label{Ber2Rog}

The formal similarity of \eqref{RogExpans} to \eqref{SupFctExp} suggests
the existence of a connection
between both approaches. Indeed, it is possible to assign to any Berezin
smf a Rogers $H^\infty$ smf which, however, will depend also on the
choice of the Grassmann algebra of constants $B$.

While \cite{[Rog]} gave a chart-by chart construction, we give here,
following \cite{[Batch]}, a more intrinsic
mechanism. It will show that the Berezin smfs are the
fundamental ones; moreover, it will be in some sense prototypical for
the connection of the naive notion of configurations to our one.

We first recall some well-known higher nonsense from
category theory which today has become a basic tool in algebraic geometry.
Given a category $\CatC$, every object $X$ generates a
cofunctor (=contravariant functor)
\Beq
 X(\cdot): \CatC\to\Sets,\quad Z\mapsto X(Z)
\Eeq
where $X(Z)$ is simply the set of all morphisms from $Z$ to $X$. Also,
every morphism $X\to X'$ generates a natural transformation
$X(\cdot)\to X'(\cdot)$. Thus, we get a functor
\Beqn X2X(cdot)
 \CatC\to\{\text{category of cofunctors $\CatC\to\Sets$}\},\quad
 X\mapsto X(\cdot),
\Eeq
and it is a remarkable observation that this functor is faithfully full,
that is, every natural transformation $X(\cdot)\to X'(\cdot)$ is generated
by a unique morphism $X\to X'$.

Moreover, let be given some cofunctor $F:\CatC\to\Sets$.
If there exists an object $X$ of $\CatC$ such that $F$ is isomorphic
with $X(\cdot)$ then this object is uniquely determined up to isomorphism;
thus, one can use cofunctors to characterize objects.

In algebraic geometry, one often calls the elements of $X(Z)$ the
{\em $Z$-valued points of $X$}. Indeed, an algebraic manifold $X$ usually
encodes a system of equations in affine or projective space which
cut it out, and if $Z$ happens to be the spectrum of a ring,
$Z=\Spec(R)$, then $X(Z)$ is just the set of solutions of this
system with values in $R$. One also writes simply $X(R)$
for this.

Now fix a finite-dimensional Grassmann algebra $B=\Lambda_k$. This is
just the algebra of global superfunctions on the linear smf
$\L(\mathbb R^{0|k})$; the underlying manifold is simply a point.
$\L(\mathbb R^{0|k})$ should be viewed
as $C^\infty$ super variant of $\Spec(B)$, and therefore it is natural to
denote, for any Berezin-Leites-Kostant smf $X$, by $X(B)$ the set of all
smf morphisms $\L(\mathbb R^{0|k})\to X$.

The set $X(B)$ will be the underlying set of the $H^\infty$ smf we are
going to construct. In fact, using Thm. \ref{FinDimCoords},
any such morphism is uniquely characterized by the pullback of global
superfunctions; thus, $X(B)$ identifies with the set of all
algebra homomorphisms $\Oinf(X)\to B$.

Given a morphism $f: X\to Y$ of Berezin smfs, we get a map
$f:X(B)\to Y(B)$, \ $u\mapsto u\circ f^*$. In particular, for open
$U\seq X$, we get an injective map $U(B)\to X(B)$ which we will view as
inclusion. The $U(B)$ are the open sets of a non-Hausdorff topology which
is just a global variant of the deWitt topology mentioned above.

Now fix a superchart on $X$, i. e. an isomorphism $U \too\phi U'$ where $U$
is open in $X$ and $U'$ is open in $\L(\mathbb R^{m|n})_{x|\xi}$.
We get a map $U(B) \too\phi U'(B)$; the target of this map
identifies naturally with a "Rogers superdomain":
\Beq
 U'(B) \cong \epsilon^{-1}(U')\seq B\seven^m\times B\sodd^n,\qquad
 f\mapsto (f(x_1),\dots,f(x_m),f(\xi_1),\dots,f(\xi_n)).
\Eeq
Denote the composite $U(B) \too\phi U'(B)\to \epsilon^{-1}(U')$
by $c_\phi$; it will become a superchart on $X(B)$.

Equip $X(B)$ with the strongest topology such that all $c_\phi$ arising from
all possible supercharts on $X$ are continuous (of course, an atlas is
sufficient, too). It follows that, since the transition map between any
two such supercharts is $H^\infty$, these supercharts equip
$X(B)$ with the structure of a $H^\infty$ smf, and thus, a fortiori, a
$G^\infty$ smf.

Also, the epimorphism \eqref{TheRestEpi} globalizes:
For $f\in\Oinf(U)$ with open $U\seq X$, we get a map $f':U(B)\to B$,
\ $u\mapsto u(f)$, which is a $H^\infty$ superfunction.
One gets an algebra epimorphism $\Oinf(U)\to H^\infty(U(B))$ which
is an isomorphism iff $n\ge k$.

Finally, given a morphism $f: X\to Y$ of Berezin smfs, the map
$f:X(B)\to Y(B)$ considered above is of class $H^\infty$.

Altogether, we get for fixed $B$ a functor
\Beqn Ber2RogFun
 \BerSmf=\{\text{category of Berezin smfs}\} \to
 \{\text{category of $H^\infty$ smfs}\},\quad X\mapsto X(B).
\Eeq
In particular, any ordinary smooth manifold $X$ can be viewed as a
Berezin smf and therefore gives rise to a $H^\infty$ smf $X(B)$. Roughly
speaking, while $X$ is glued together from open pieces of $\mathbb R^m$ with
smooth transition maps, $X(B)$ arises by replacing the open piece
$U\seq\mathbb R^m$ by $\epsilon^{-1}(U)\seq B\seven^m$, and the transition maps
by their Grassmann analytic continuations.

\Brm
(1) Apart from the fact that $B$ is supercommutative instead of commutative,
the functor \eqref{Ber2RogFun} is just the Weil functor
considered in \cite{[KoMiSl]}.

(2)
Looking at the $Z$-valued points of an smf is often a useful technique
even if one stays in the Berezin framework. For instance, a Lie supergroup
$G$, i.~e. a group object in the category of supermanifolds, turns in this
way to a functor with values in the category of ordinary groups
(by the way, this property is not shared by quantum groups, which makes them
much more difficult objects). While, using
the diagrammatic definition of a group object, it is by no means trivial that
the square of the inversion morphism $\iota: G\to G$ is the identity,
this fact becomes an obvious corollary of the corresponding property
of ordinary groups by using $Z$-valued points.
Cf. \cite{[SchReport]}, \cite{[Lei1]}.
\Erm

\subsection{Molotkov's approach and the r\^ole of the algebra of constants}

For a review of Molot\-kov's approach \cite{[Mol 84]} to
infinite-dimensional supermanifolds, and a comparison with our one,
cf. \cite{[IS2]}. Here we give an interpretation of the idea in
categorial terms.

Roughly said, this approach relies on the description of smfs by
their points with values in $0|k$-dimensional smfs,
where, however, all $k$'s are considered simultaneously.

The basic observation to understand Molotkov's approach is the following:
Let $\BerSmf$ be the category of all Berezin smfs, and let
$\Gr$ be the category the objects of which are the real Grassman
algebras $\Lambda_k=\Lambda[\xi_1,\dots,\xi_k]$
(one for each $k\ge0$), with all even algebra homomorphisms as morphisms.
Assigning to every $\Lambda_k$ the $0|k$-dimensional smf
$\Spec(B):=\L(\mathbb R^{0|k})=(\text{point},\Lambda_k)$ we get a cofunctor
$\Spec: \Gr\to\BerSmf$ which establishes an equivalence of the
opposite category $\Gr^{\opn Op }$ with
the full subcategory $\Spec(\Gr)$ of $\BerSmf$ of those smfs
which are connected and have even dimension zero.

On the other hand, it is easy to see that the subcategory $\Spec(\Gr)$
is sufficient to separate morphisms, that is, given two distinct smf morphisms
$X\doublearrow X'$ there exists some $k>0$ and a morphism
$\Spec(\Lambda_k)\to X$ so that the composites
$\Spec(\Lambda_k)\to X\doublearrow X'$ are still different.

Thus, if assigning to any smf $X$ the composite functor
\Beq
 \Gr \too\Spec  \BerSmf \too{X(\cdot)} \Sets,
\Eeq
we get a functor
\Beqn Smf2Functs
 \BerSmf\to\{\text{category of functors $\Gr\to\Sets$}\},\quad
 X\mapsto X(\cdot),
\Eeq
and, due to the separation property mentioned above, this functor is
still faithful, i. e. injective on morphisms. (It is unlikely that it is full,
but I do not know a counterexample.) Thus, an smf is characterized
up to isomorphism by the functor $\Gr\to\Sets$ it generates.

However, there exist functors not representable by smfs. Roughly said,
a given functor $F: \Gr\to\Sets$ is generated by an smf iff it can be
covered (in an obvious sense) by "supercharts", i. e. subfunctors which
are represented by superdomains. However, as a consequence of the lack
of fullness of the functor \eqref{Smf2Functs}, we cannot guarantee a
priori that the transition between two supercharts is induced by a
morphism of superdomains -- we simply have to require that.

Explicitly, given a linear smf $\L(E)$ where $E\cong\mathbb R^{m|n}$
is a finite-dimensional \ztwo-graded vector space, one has for each
$B:=\Lambda_k$
\Beq 
 \L(E)(B) = (B\otimes E)\seven \cong B\seven^m\times B\sodd^n;
\Eeq
we encountered this already in the previous section.

Now, Molotkov's idea is to replace here $E$ by a Banach space and
define just the linear smf generated by $E$ as the functor
$F_E: \Gr\to\Sets$,\ $B\mapsto(B\otimes E)\seven$. He then defines
a "superregion" alias superdomain as an in a suitable sense open subfunctor of
this, and an smf as a functor $F: \Gr\to\Sets$  together with an
atlas of subfunctors such that the transition between two of them
is supersmooth in a suitable sense.

Actually, his functors have their values not in the set category but in the
category of smooth Banach manifolds, but this is a structure which cares for
itself, thanks to the use of atlasses.

Resuming we find that
\begin{itemize}
\item
the Berezin-Kostant approach works without any "algebra of constants" except
$\mathbb R$ or $\mathbb C$, i. e. without an auxiliary Grassmann algebra;
\item
the deWitt-Rogers approach works with a fixed Grassmann or Grassmann-like
algebra of constants;
\item
Molotkov uses all finite-dimensional Grassmann algebras at the same
time as algebras of constants, and thus achieves independence of a
particular choice.
\end{itemize}

My personal point of view is that any differential-geometric concept
which makes explicit use of the algebra of constants $B$, and hence is not
functorial w. r. to a change of $B$, is suspicious.
Although the possibility of choice of a concrete $B$ allows
new differential-geometric structures and phenomena, which
by some people are claimed to be physically interesting,
there is up to now no convincing argument for their potential
physical relevance.

In particular, this is connected with the fact that there exist
Rogers smfs which do not arise from a Berezin smf by the construction
given above. A prototypical example is a $1|1$-dimensional supertorus,
in which also the odd coordinate is wrapped around. It is clearly
non-functorial in $B$ because it needs the distinction of a discrete
subgroup in $B\sodd$.

Also, if one defines $G^\infty$ Lie supergroups in the obvious way, the
associated Lie superalgebra will be a module over $B$, with the
bracket being $B$-linear. It follows that the definition is bound to
a particular $B$.

In particular, if $G$ is a Lie supergroup in Berezin's sense, with
Lie superalgebra $\mathfrak g$, then, for a Grassmann algebra $B$, the
Rogers Lie supergroup $G(B)$ will have Lie superalgebra ${\mathfrak g}\otimes
B$;
Now it is an abundance of Lie superalgebras over $B$ which are not of
this form; cf. \cite[4.1]{[DeW2]} for an example of a $0|1$-dimensional
simple Lie superalgebra which cannot be represented as ${\mathfrak g}\otimes
B$.
However, none of these "unconventional" Lie superalgebras seems to
be physically relevant.

A general metamathematical principle, which is up to now supported by
experience, is: Any mathematical structure which does not make
{\em explicit} use of the algebra of constants $B$ can be (and should
be) formulated also in Berezin terms.

In particular, this applies to the structures appearing in geometric
models in quantum field theory.

\subsection{Comparison of the approaches}\label{CompOfAppr}

The deWitt-Rogers approach has quite a few psychological advantages:
Superfunctions are now $B$-valued maps, i. e. genuine functions on this
space instead of elements of an abstract algebra. Also, instead of
morphisms of ringed spaces one has genuine superdifferentiable
maps.

Also, from an aesthetic point of view, even and odd degrees of freedom
seem to stand much more on equal footing: While the underlying space of a
Berezin smf encodes only even degrees of freedom, the underlying space
of a Rogers-like smf is a direct product of even and odd part.

Finally, in the Taylor expansion \eqref{RogExpans} of a $G^\infty$
function at $\xi=0$, the functions $z(f_\mu)(u)$ are still $B$-valued,
and thus they may perfectly well anticommute (they do so iff
they are $B\sodd$-valued). At the first glance, this approach appears to
be ideally suited for the implementation of the anticommutativity
of fermionic field components.

These appealing features tempted quite a few physicists to prefer
this approach to the Berezin approach (cf. in particular \cite{[DeW2]}).
But one has to pay for them:
"This description, however, is extremely
uneconomical because in the analysis do the $\xi$'s" (here $\theta$'s)
"explicitly enter. So, the $\xi$'s are really unnecessary"
(\cite[6.5.3]{[Hen/Tei]}).

For instance, the $4|4$-dimensional
super Minkowski space, that is, the space-time of e.g. $N=1$
super Yang-Mills theory with superfields, is modelled as
$X=B\seven^4\times B\sodd^4$. Thus, one has an "inflation of points", and
even the most eager advocates of this approach have no idea how to
distinguish all these points physically. In fact, the algebra $B$ plays
an auxiliary role only. Notwithstanding which of the modern,
mainly heuristic, concepts of quantization of a field theory
one uses, $B$ does not appear at all at the quantized level.
In other words, $B$ is an unphysical "addendum" used to formulate
the classical model. This becomes even clearer if we try to
model fermion fields on the ordinary Minkowski space within this
approach: Space-time would be described by $B\seven^4$ instead of
the usual (and certainly more appropriate) $\mathbb R^4$.

On the other hand, one can obtain anticommuting field components
necessary for modelling fermion fields also in the framework
of Berezin's approach: Instead of real- or complex-valued functions
or superfunctions $f\in\Oinf(X)$ one uses "$B$-valued" ones:
$f\in\Oinf(X)\otimes_{\mathbb R} B$ where $B$ is a Grassmann algebra or
some other suitable \ztwo-commutative algebra, and the tensor
product is to be completed in a suitable sense if $B$ is
infinite-dimensional.

This removes the inflation of points -- but the unphysical auxiliary
algebra $B$ remains. At any rate, this argument shows that the deWitt-Rogers
approach has not the principal superiority over Berezin's approach
which is sometimes claimed by its advocates.
The problem of "$B$-valued configurations" will be discussed in the
next section.

The question arises, what algebra of constants $B$ should we take,
how "large" should it be? As we saw in \ref{Sketch}, in particular, in
considering the epimorphism \eqref{TheEpi}, a finite-dimensional
$B$ leads to various unwanted truncation effects; these can be avoided
by taking $B$ infinite-dimensional. But even this decision leaves still room;
instead of a Banach completion of a Grassmann algebra with a countable
set of generators, one could use e.g. one of the Banach algebras
${\mathcal P}(p;\mathbb R)$ introduced in \cite{[CMP1]} (with a purely odd
field
target space).

On the other hand, an $m|n$-dimensional smf will now become as topological
space $m\cdot\dim(B\seven)+n\cdot\dim(B\sodd)$-dimensional. One
feels uneasy with such an inflation of dimensions. Moreover, we cannot use
the manifolds of ordinary differential geometry directly -- we have to
associate to them the corresponding smfs (although it is an almost
everywhere trivial process to "tensor through" ordinary structures
to their $B$-analogues, for instance, this applies to metrics, connections,
vector bundles etc).

Still worse, the automorphism group $\opn Aut _{\mathbb R}(B)$, which is at
least
for Grassmann algebras pretty big, appears as global symmetry
group; thus, we have {\em spurious symmetries}. Indeed, given a
Lagrange density which is a differential polynomial
${\mathcal L}\in\mathbb C\br<\ul\Phi|\ul\Psi>$ and an automorphism $\alpha:
B\to B$,
the field equations will stay invariant under
$(\Phi,\Psi)\mapsto(\alpha\Phi,\alpha\Psi)$. (This is analogous to the
process in which the only continuous automorphism of $\mathbb C$, the
conjugation,
induces the charge conjugation for a complex field.)
Of course, this symmetry could be inhibited by using a
differential polynomial with coefficients in $B$, i.~e.
${\mathcal L}\in B\br<\ul\Phi|\ul\Psi>$, but all usual models have coefficients
in $\mathbb C$.

One feels that the algebra of constants $B$ is an addendum, something which
is not really a part of the structure, and that it should be thrown out.
This impression is still stronger if we look at a quantized theory:
Sometimes, for instance in \cite{[DeW2]}, it is claimed that
the result of quantization should be a a "super Hilbert space", i. e.
a $B$-module ${\mathcal H}$ together with a skew-linear $B$-valued scalar
product
$\br<\cdot|\cdot>$ such that the body of $\br<\phi|\phi>$ for
$\phi\in{\mathcal H}$ is non-negative, and is positive iff the body of $\phi$
is
non-vanishing.

However, no one has ever measured a Grassmann number, everyone measures
real numbers. Moreover, while a bosonic field strength operator
$\widehat\Phi(x)$ (or, more exactly, the smeared variant
$\int dx\;g(x)\widehat\Phi(x)$) encodes in principle an observable
(apart from gauge fields, where only gauge-invariant expressions are
observable), a fermionic field strength operator $\widehat\Psi(x)$
(or $\int dx\;g(x)\widehat\Psi(x)$) is only a "building block" for
observables. That is, the eigenvalues of
$\int dx\;g(x)\widehat\Psi(x)$ do not have a physical meaning.

So it is in accordance with the quantized picture to assume on the classical
level that bosonic degrees of freedom are real-valued while the fermionic
ones do not take constant values at all. (In fact, the picture is somewhat
disturbed by the fact that, in the quantized picture, multilinear combinations
of an even number of fermion fields do take values, like e. g. the electron
current, or the pion field strength; that is, although classical fermion
fields have only "infinitesimal" geometry, their quantizations generate
non-trivial, non-infinitesimal geometry.)

\subsection{$B$-valued configurations}

Suppose we have a classical field model on $\mathbb R^4$. Usually, the theory
is described by saying what boson fields $\Phi(\cdot)$ and fermion fields
$\Psi(\cdot)$ appear, and a Lagrange density ${\mathcal L}[\Phi,\Psi]$, which,
at the first glance, is a functional of these fields.

In the language of \cite{[CMP1]}, the model is given by fixing a field
target space $V$, which is a finite-dimensional \ztwo-graded vector
space on which the two-fold cover of the Lorentz group acts, and a
differential polynomial
${\mathcal L}[\ul\Phi,\ul\Psi]\in \mathbb C\br<\ul\Phi|\ul\Psi>$.

The most obvious way to implement \eqref{AntiComm} is to assume that the
fermion fields $\Psi(\cdot)$ are functions on $\mathbb R^4$ with values
in the odd part of a \ztwo-graded commutative algebra $B$, for
instance, a Grassmann algebra $\Lambda_n$.

Now, if one takes the field equations serious (indeed, they should give the
first order approximation of the quantum dynamics) then it follows that
boson fields cannot stay any more real- or complex-valued; instead, they
should have values in the even part of $B$. Cf. e. g. \cite{[Choquet-Bruhat]}
for a discussion of the classical field equations of supergravity
within such an approach.

Also, if the model under consideration works with superfields, then these
should be implemented as $B$-valued superfunctions. For instance, the
chiral superfield \eqref{ChirSupF} is now implemented as an element
$\Phi\in\left(\Oinf(\L(\mathbb R^{4|4}))\otimes B\right)\seven$ which satisfies
the constraint $\cj D_{\dot\alpha} \Phi = 0$; the requirement that $\Phi$ be
even leads to the required anticommutativity of the Weyl spinor
$\psi\in C^\infty(\mathbb R^4,\mathbb C^2)\otimes B\sodd$.

Now there are arguments which indicate that the
"$B$-valued configurations" should not be the ultimate solution of the
problem of modelling fermion fields: First, as in \ref{CompOfAppr},
it is again not clear which algebra of constants $B$ one should take.

Also, any bosonic field component
will have $\dim(B\seven)$ real
degrees of freedom instead of the expected single one. Thus,
we have {\em fake degrees of freedom}, quite analogous to the
"inflation of points" observed in \ref{CompOfAppr}.

Our remarks in \ref{CompOfAppr} on {\em spurious symmetries} as well
as on the inobservability of Grassmann numbers apply also here.

Also, the $B\seven$-valuedness of the bosonic fields
makes their geometric interpretation problematic.
Of course, it is still possible to interpret a $B\seven$-valued
gauge field as $B\seven$-linear connection in a bundle of $B\seven$-modules
over $\mathbb R^4$ (instead of an ordinary vector bundle, as it
is common use in instanton theory). But what is the geometric
meaning of a $B\seven$-valued metric $g_{\mu\nu}$?

For every finite-dimensional $B$ there exists
some $N$ such that the product of any $N$ odd elements vanishes, and this
has as consequence the unphysical {\em fake relation}
\Beq
 \Psi_{j_1}(x_1) \cdots \Psi_{j_N}(x_N) =0.
\Eeq
This fake relation can be eliminated by using an infinite Grassmann algebra,
like e. g. the deWitt algebra $\Lambda_\infty$, or some completion of it; but
the price to be paid is that we now use infinitely many real degrees of
freedom in order to describe just one physical d. o. f.

Finally, in the functional integral
\Beq
 \bigl<\hat A\bigr>_{\opn vac } =
 \frac 1N \int [D\Phi][D\cj\Psi][D\Psi]\exp(\i S[\Phi,\cj\Psi,\Psi]/\hbar)
 A[\Phi,\cj\Psi,\Psi]
\Eeq
-- however mathematically ill-defined it may be -- the bosonic
measure $[D\Phi]$ still runs (or is thought to do so) over
$\mathbb R$-valued configurations, not over $B\seven$-valued ones.

All these arguments indicate that $B$-valued configurations should not be the
ultimate solution to the problem of modelling fermionic degrees of freedom.
Nevertheless, we will see in a moment that they do have a satisfactory
geometric interpretation, at least if $B=\Lambda_k$ is a finite-dimensional
Grassmann algebra: They should be thought of as {\em families of
configurations} parametrized by the $0|k$-dimensional smf $\L(\mathbb
R^{0|k})$.

\subsection{Families of configurations}\label{FamPhilo}
Historically, the "family philosophy" comes from algebraic geometry;
it provides a language for the consideration of objects varying
algebraically (or analytically) with parameters. It turns out to be
rather useful in supergeometry, too; cf. \cite{[Lei1]}, \cite{[SchReport]},
\cite{[IS2]}.

We begin with the standard scalar field theory given by the Lagrangian density
\Beq
 {\mathcal L}[\ul\Phi] =\sum_{a=0}^3 \partial_a\ul\Phi\partial^a\ul\Phi
 + m^2\ul\Phi^2 + V(\ul\Phi)
\Eeq
where $V$ is a polynomial, e. g. $V(\ul\Phi) = \ul\Phi^4$.

Thus, a configuration is simply a real function on $\mathbb R^4$; for
convenience,
we will use smooth functions, and the space of all configurations is the
locally convex space $M=C^\infty(\mathbb R^4)$. For any bounded open region
$\Omega\seq\mathbb R^4$, the action on $\Omega$ is a real-analytic function
\Beqn BosAction
 S_\Omega[\cdot]: M\to\mathbb R,\quad\phi\mapsto
 S_\Omega[\phi]:=\int_\Omega d^4x{\mathcal L}[\phi].
\Eeq

Now suppose that $\phi=\phi(x_0,\dots,x_3|\zeta_1,\dots,\zeta_n)$ depends
not only on the space-time variables $x_0,\dots,x_3$ but additionally on
odd, anticommuting parameters $\zeta_1,\dots,\zeta_n$. In other words,
$\phi\in\Oinf(\mathbb R^4\times\L(\mathbb R^{0|n}))\sevR$ is now a
{\em family of configurations} parametrized by the $0|n$-dimensional smf
$Z_n:=\L(\mathbb R^{0|n})$ (we require $\phi$ to be even in order to
have it commute).

As in \eqref{SupFctExp} we may expand
$\phi(x|\zeta)=
 \sum_{\mu\in\mathbb Z_2^n,\ \br|\mu|\equiv0(2)} \phi_\mu(x)\zeta^\mu$,
i. e. we may view $\phi$ as a smooth map
\Beq
 \phi: \mathbb R^4 \to \mathbb C[\zeta_1,\dots,\zeta_n]\sevR.
\Eeq
That is, $\phi$ is nothing but a $B$-valued configuration with
the Grassmann algebra $B:= \mathbb C[\zeta_1,\dots,\zeta_n]$ as
auxiliary algebra!

Encouraged by this, we look at a model with fermion fields, say
\Beqn BDrDismant
 {\mathcal L}[\ul\Phi|\ul\Psi] := \frac \i2 \sum_{a=0}^4 \br(
  \dcj{\ul\Psi}\gamma^a\partial_a\ul\Psi
  - \partial_a\dcj{\ul\Psi}\gamma^a\ul\Psi)
  - m^\Psi\dcj{\ul\Psi}\ul\Psi
  - \sum_{a=0}^4 \partial_a\ul\Phi\partial^a\ul\Phi - (m^\Phi)^2\ul\Phi^2
  - \i g \dcj{\ul\Psi}\ul\Psi\ul\Phi
\Eeq
where $\ul\Phi$ is a scalar field, $\ul\Psi$ a Dirac spinor, and
the thick bar denotes the Dirac conjugate (in fact, \eqref{BDrDismant}
is a slightly dismantled version of the Yukawa model of
meson-nucleon scattering). 

Note that \eqref{BDrDismant} is well-defined as a differential polynomial
${\mathcal L}[\ul\Phi|\ul\Psi]\in\mathbb C\br<\ul\Phi|\ul\Psi>\sevR$.

While a naive configuration for the fermion field,
$\psi=(\psi_\alpha) \in C^\infty(\mathbb R^4,\mathbb C^4)$, would violate the
requirement of anticommutativity of the field components $\psi_\alpha$, this
requirement is satisfied if we look for a family
$(\psi_\alpha) = (\psi_\alpha(x_0,\dots,x_3|\zeta_1,\dots,\zeta_n))$, which now
consists of odd superfunctions
$\psi_\alpha\in\Oinf(\mathbb R^4\times\L(\mathbb R^{0|n}))\sodd$. Looking
again at the expansion, we can reinterpret $\psi_\alpha$ as a
smooth map
\Beq
 \psi_\alpha: \mathbb R^4 \to \mathbb C[\zeta_1,\dots,\zeta_n]\sodd,
\Eeq
that is, $\psi$ is nothing but a $B$-valued configuration again.

Resuming, we can state that if $B= \mathbb C[\zeta_1,\dots,\zeta_n]$ is
a Grassmann algebra then {\em $B$-valued configurations can be viewed as
families of configurations parametrized by the smf $\L(\mathbb R^{0|n})$.}

Now there is no need to use only $0|n$-dimensional smfs as parameter space.
Thus, for the model \eqref{BDrDismant}, a {\em $Z$-family of configurations}
where $Z$ is now an arbitrary Berezin supermanifold, is a tuple
\Beqn TheFam
 (\phi,\psi_1,\dots,\psi_4)\in
 \Oinf(\mathbb R^4\times Z)\sevR\times
 \prod^4 \Oinf(\mathbb R^4\times Z)\sodd,
\Eeq
and the action over any compact space-time domain
$\Omega\seq\mathbb R^4$ becomes an element
\Beqn FamAction
 S_\Omega[\phi|\psi]:=\int_\Omega d^4x{\mathcal L}[\phi|\psi]\in
 \Oinf(Z)\sevR.
\Eeq
Also, if $\mu:Z'\to Z$ is some smf morphism then \eqref{TheFam} can be
pullbacked along the morphism
$1_{\mathbb R^4}\times\mu :\mathbb R^4\times Z'\to \mathbb R^4\times Z$ to give
a
$Z'$-family $(\phi',\psi'_1,\dots,\psi'_4)$.

The simplest case is that $Z=P$ is a point: Since
$\Oinf(\mathbb R^4\times P)=C^\infty(\mathbb R^4)$, every $P$-family has the
form
$(\phi|0,0,0,0)$ where $\phi\in C^\infty(\mathbb R^4)$.
Thus it encodes simply a configuration for the bosonic sector.

More generally, if $Z$ has odd dimension zero, i. e. is essentially an
ordinary manifold, then for $Z$-family we have $\psi_\alpha=0$ by evenness.
Hence {\em we need genuine supermanifolds as parameter spaces in order
to describe non-trivial configurations in the fermionic sector}.

If $Z$ is a superdomain with coordinates $z_1,\dots,z_m|\zeta_1,\dots,\zeta_n$
then \eqref{TheFam} is simply a collection of one even and four odd
superfunctions which all depend on
$(x_0,\dots,x_3,z_1,\dots,z_m|\zeta_1,\dots,\zeta_n)$, and the pullbacked
family now arises simply by substituting the coordinates
$(z_i|\zeta_j)$ by their
pullbacks in $\Oinf(Z')$; this process is what a physicist would call a
{\em change of parametrization}.

This suggests to look for a {\em universal family} from which every other
family arises as pullback. This universal family would then encode all
information on classical configurations.

Of course, the parameter smf $M_{C^\infty}$ of such a family, i. e.
the {\em moduli space} for configuration families, is necessarily
infinite-dimensional;
hence we will have to extend the Berezin-Leites-Kostant calculus (or,
strictly spoken, its hermitian variant as sketched in \ref{HermStuff})
to a calculus allowing \ztwo-graded locally convex spaces as model spaces;
we will do so in section \ref{InfDSmfs}.

So, sticking to our example Lagrangian \eqref{BDrDismant},
we are looking for an $\infty|\infty$-dimensional smf $M_{C^\infty}$
and a tuple of smooth superfunctions
\Beqn TheUnivFam
 (\Phi,\Psi_1,\dots,\Psi_4)\in
 \Oinf(\mathbb R^4\times M_{C^\infty})\sevR\times
 \prod^4 \Oinf(\mathbb R^4\times M_{C^\infty})\sodd
\Eeq
such that for any other family \eqref{TheFam} there exists a unique
morphism of smfs $\mu_{(\phi,\psi_1,\dots,\psi_4)}:Z\to M$
such that \eqref{TheFam} is the pullback
of \eqref{TheUnivFam} along this morphism. We call
$\mu_{(\phi,\psi_1,\dots,\psi_4)}$ the {\em classifying} morphism
of the family \eqref{TheFam}.

However, we will not implement \eqref{TheUnivFam} in the verbal sense
since a definition of {\em smooth} superfunctions in the
infinite-dimensional case, which is needed to give the symbol
"$\Oinf$"
in \eqref{TheUnivFam} sense, is technically rather difficult, and we will
save much work by sticking to real-analytic infinite-dimensional
supermanifolds.

Fortunately, this difficulty is easily circumvented: We simply consider
only those families \eqref{TheFam} for which $Z$ is actually a real-analytic
smf, and the $Z$-dependence is real-analytic (note that if
$Z=\L(\mathbb R^{0|n})$ is purely even then this requirement is empty since
we have actually polynomial dependence). Once the sheaf
$\O^{C^\infty(\mathbb R^4)}(\cdot)$ of superfunctions on $Z$ with values in
the locally convex space $C^\infty(\mathbb R^4)$ has been defined
(cf.  \ref{SFcts} below), we can then rewrite \eqref{TheFam} to an element
\Beq
 (\phi,\psi_1,\dots,\psi_4)\in
 \O^{C^\infty(\mathbb R^4)}(Z)\sevR\times
 \prod^4 \O^{C^\infty(\mathbb R^4)}(Z)\sodd.
\Eeq
Comparing with Cor. \ref{FinDimPbChar} and its infinite-dimensional version
Thm. \ref{CoordChar} below it is now easy to see how to construct the
universal family: Its parameter smf is simply a linear smf,
$M_{C^\infty} = \L(E)$, with the \ztwo-graded locally convex space
\Beq
 E = \underbrace{C^\infty(\mathbb R^4)}_{\text{even part}}\ \oplus\
    \underbrace{C^\infty(\mathbb R^4)\otimes \mathbb C^{0|4}}_{\text{odd part}}
\Eeq
as model space, and the universal family \eqref{TheUnivFam}
is now just the standard coordinate on this infinite-dimensional smf:
\Beq
 (\Phi,\Psi_1,\dots,\Psi_4)\in
 \O^{C^\infty(\mathbb R^4)}(M_{C^\infty})\sevR\times
 \prod^4 \O^{C^\infty(\mathbb R^4)}(M_{C^\infty})\sodd.
\Eeq
Comparing with \ref{Ber2Rog}, we see that the set of $B$-valued
configurations stands in the same relation to our configuration
smf $M_{C^\infty}$ as a finite-dimensional Berezin smf $X$ to the
associated $H^\infty$-deWitt smf $X(B)$.
That is, all information is contained in $M_{C^\infty}$, which therefore
should be treated as the fundamental object.

For instance, as special case of \eqref{FamAction} and as
supervariant of \eqref{BosAction},
the action on a bounded open region $\Omega\seq\mathbb R^4$
is now a real-analytic scalar superfunction on $M$:
\Beqn UnivAction
 S_\Omega[\Phi|\Psi]:=\int_\Omega d^4x{\mathcal L}[\Phi|\Psi]
\in\O(M_{C^\infty})\sevR,
\Eeq
and the action \eqref{FamAction} of any family $(\phi|\psi)$ is just
the pullback of \eqref{UnivAction} along its classifying morphism.
Thus, the element \eqref{UnivAction} can be called the
{\em universal action} on $\Omega$.

Also, the field strengthes at a space-time point $x\in\mathbb R^4$
become superfunctions:
\Beq
 \Phi(x)\in\O(M_{C^\infty})\sevR,\quad
 \Psi_\alpha(x)\in\O(M_{C^\infty})\sodd.
\Eeq
It is now the \ztwo-graded commutativity of $\O(M_{C^\infty})$ which ensures
the validity of \eqref{AntiComm}.

\subsection{Classical configurations, functionals of classical
fields, and supermanifolds}\label{Program}

Here we formulate our approach programmatically; an exposition of
bourbakistic rigour will follow in the successor papers \cite{[CMP3]},
\cite{[CMP4]}.

\subsubsection{Classical configurations}
The keystone of our approach is the following:

{\em The configuration space of a classical field model with fermion
fields is an $\infty|\infty$-dimensional supermanifold
$M$. Its underlying manifold is the set $M_{\opn bos }$ of all configurations
of the bosonic fields while the fermionic degrees of freedom
are encoded as the odd dimensions into the structure sheaf $\O_M$.

The functionals of classical fields, which we described in
\cite{[CMP1]} as superfunctionals, are just the superfunctions on $M$.
If such a functional describes an observable it is necessarily even
and real.}

Suppose that our model describes $V$-valued fields on flat space-time
$\rdmm$, as in \cite{[CMP1]} (this is the case for almost all common
models in Minkowski space; in Yang-Mills theory, it is the case after
choosing a reference connection; however, it is no longer true for
$\sigma$ models or for models including gravitation in the usual way).
In that case, the "naive configuration space" is a suitable (cf. Rem. (1)
below) admissible space $E$ with respect to the setup $(d,V)$ (we recall that
this means ${\mathcal D}(\rdmm)\otimes V \seq E \seq {\mathcal
D}'(\rdmm)\otimes V$
with dense inclusions), and the supermanifold of configurations is just
the linear smf
\Beq
 \L(E):=(E\seven,\O_E),
\Eeq
i. e. it has underlying space $E\seven$, while the structure sheaf is
formed by the superfunctionals introduced in \cite{[CMP1]} (cf. \ref{Smfs}
below for the precise definition of smfs). For more general models,
e. g. $\sigma$ models, the configuration space has to be glued together
from open pieces of linear superdomains.

The flaw of the usual attempts to model fermion
fields lies in the implicite assumption that the configuration
space $M$ should be a {\em set}, so that one can ask for the
form of its elements. However, if $M$ is a supermanifold in
Berezin's sense then it has no elements besides that of its underlying
manifold $M_{\opn bos }$, which just correspond to configurations
with all fermion fields put to zero.

However, although "individual configurations" do not exist (besides
the purely bosonic ones), {\em families of configurations} parametrized
by supermanifolds $Z$ do exist. Precise definitions will be given
in the successor papers. Here we note:

{\em A $Z$-family of configurations encodes the same information as a
morphism of supermanifolds $Z\to M$.}

Thus, $M$ can be understood as the representing object for the
cofunctor
\Beq
 \Smfs\to \Sets,\quad
 Z\mapsto\{\text{$Z$-families of configurations}\}.
\Eeq
In particular, let us look for field configurations with values in a
finite-dimensional complex Grassmann algebra
$\Lambda_n:=\mathbb C[\zeta_1,\dots,\zeta_n]$.
Since $\Lambda_n=\O(\L(\mathbb R^{0|n}))$,
such a $\Lambda_n$-valued configuration should be viewed
as $\L(\mathbb R^{0|n})$-family of configurations,
and thus encodes a morphism of supermanifolds $\L(\mathbb R^{0|n})\to M$.

Thus, the set of all $\Lambda_n$-valued configurations is in natural bijection
with the set of all morphisms from $\L(\mathbb R^{0|n})$ to $M$. This shows
that the really fundamental object is $M$, and it explains the
arbitrariness in the choice of the "auxiliary algebra" $B$ in any Rogers-like
formulation.

With this point of view, we can do without Rogers' supermanifolds
at all. Even better, one has a functor from "Berezin things"
to "Rogers things": given a Berezin supermanifold $X$ the set
of all $\L(\mathbb R^{0|n})$-valued points of $X$ forms in a natural way
a Rogers supermanifold.

\Brm
(1) The precise meaning of the notion
"configuration space" depends on the choice of the model space
$E$, i. e. on the functional-analytic quality of the configurations
to be allowed. If $E$ is too large (for instance, the maximal
choice is $E={\mathcal D}'(\rdmm)\otimes V)$ then we have few classical
functionals (in our example, all coefficient functions have to
be smooth, and there are no local functionals besides linear
ones), and the field equations may be ill-defined. On the other hand,
if $E$ is too small then we have a lot of
classical functionals but possibly few classical solutions of
the field equations; e.g., for $E={\mathcal D}(\rdmm)\otimes V$,
there are no non-vanishing classical solutions.

If necessary one can consider different configuration spaces
for one model; the discussion at the end of \ref{Schwinger} suggests
that this might be appropriate.
In the case of models with linear configuration
space, each choice of $E$ defines its own configuration supermanifold
$\L(E)$. We note that a continuous inclusion of admissible spaces,
$E\seq E'$, induces a morphism of supermanifolds $\L(E)\to\L(E')$.

(2) The question arises whether the supermanifold $M$ should
be analytic, or smooth? Up to now, we stuck to the real-analytic
case; besides of the fact that, in an infinite-dimensional situation,
this is technically easier to handle, this choice is motivated by
the fact that the standard observables (like action, four-momentum,
spinor currents etc.) are integrated differential polynomials
and hence real-analytic superfunctionals, and so are also the
field equations which cut out the solution supermanifold $M\sol$
(cf. below).

Unfortunately, the action of the symmetry groups of the model
will cause difficulties, as we will discuss in the successor paper
\cite{[CMP4]}. Therefore it is not superfluous to remark that a
$C^\infty$ calculus of $\infty$-dimensional supermanifolds is possible.

Fortunately, if the supermanifold of classical solutions
(cf. \ref{ClAcClSols} below) exists at all, it will be real-analytic.

(3) Already B. DeWitt remarked in \cite{[DeW1]} that the configuration
space of a theory with fermions should be regarded as a Riemannian
supermanifold. Strangely enough, the same author wrote the book
\cite{[DeW2]} in which he elaborated a Rogers-like approach with all
details. Cf. also p. 230 of \cite{[DeW2]} in which he emphasizes
again that the configuration space is a supermanifold $C$ --
of course, in a Rogers-like sense again. Now his algebra $\Lambda_\infty$
of "supernumbers" turns out to be the algebra of global superfunctions
on the $0|\infty$-dimensional smf $\L(\Pi\mathbb R^\infty)$, and
$C$ is again the set of all $\L(\Pi\mathbb R^\infty)$-valued points of $M$.
Cf. \cite{[CMP3]} for details.

The "supermanifold of fields" $M$ is (somewhat incidentally) mentioned
also in \cite[18.1.1]{[Hen/Tei]}.

(4) Although our point of view could look crazy in the eyes
of physicists, it is in perfect agreement with their heuristic
methods to handle classical fermion fields. In particular, this
applies to the fermionic variant of the Feynman integral over
configurations: Consider the Gaussian integral
\Beqn GaussInt
 \int[D\cj\Psi][D\Psi]\exp(-\cj\Psi A\Psi)={\opn Det }A
\Eeq
where $\Psi$ is a fermion field. The usual, heuristic justification
of the setting \eqref{GaussInt} amounts in fact to an analogy conclusion
from the model case of the integral over a volume form over a
$0|n$-dimensional complex supermanifold (cf. \cite{[Ber2ndQ]} or modern
textbooks on quantum field theory.
{}From this point of view, it is very natural to think that the left-hand side
of \eqref{GaussInt} is in fact the integral of a volume form on a
$0|\infty$-dimensional supermanifold $M$.

(5) It is interesting to note that we make contact with supergeometry
whenever a classical field model contains fermion fields --
irrespective of whether supersymmetries are present or not. Perhaps,
it is not devious to view this as a "return to the sources". Indeed,
the source of supergeometry was just the anticommutativity of
classical fermion fields.

(6) Of course, the programme just being presented applies equally well to
string models since the latter can be viewed as two-dimensional field models.
On the other hand, we neglect all models with more complicated
("plecton" or "anyon") statistics; although they nowadays have
become fashionable, they have up to now not been proven to have fundamental
physical relevance for particle physics.
\Erm

In order to implement our approach also for more general models,
one needs a theory of infinitedimensional supermanifolds. In
\cite{[ISA]}, \cite{[IS1]}, \cite{[IS2]}, the present author constructed
a general theory of infinite-dimensional real or complex analytic
supermanifolds modelled over arbitrary locally convex topological
vector spaces. In fact, \cite{[CMP1]} presented a specialization of
elements of this theory to the case that the
model space is an admissible function spaces on $\mathbb R^d$.

In section \ref{InfDSmfs}, we abstract from the function space nature
of $E$, and we globalize the theory from superdomains to supermanifolds.
Thus, we will give an alternative description of the theory mentioned
above.

\subsubsection{Classical action, field equations, and classical solutions}
\label{ClAcClSols}
Turning to the action principle, we begin with a naive formulation which
is suitable for Euclidian models:

{\em The classical action $S$ is an observable; thus, $S\in\O(M)\sevR$.
Moreover, the condition $\delta S=0$ should cut out in $M$ a
sub-supermanifold $M\sol$, the {\em supermanifold of the
classical solutions}.}

Here
\Beq
 \delta S = \int_{\rdmm} dx\;\Bigl(
   \sum_{i=1}^{N^\Phi} \frac \delta{\delta\Phi_i(x)}S \ \delta\Phi_i(x) +
   \sum_{j=1}^{N^\Psi} \frac \delta{\delta\Psi_j(x)}S \ \delta\Psi_j(x)
  \Bigr) \in\Omega^1(M)
\Eeq
is the total variation (cf. 3.8), alias exterior differential, of $S$.

The precise meaning of the phrase "cutting out" will be explained
in \ref{CuttingOut}.

Note that this naive formulation of the action principle is doomed to
failure for models in Minkowski space: there is not a single
nonvanishing solution of the free field equations for which the
action over the whole space-time is finite. Thus, there is no element of
$\O(M)\sevR$ which describes the action over the whole space-time, and
one has to use a "localized" variant of the action principle:
Usually, the Lagrangian is given as a differential polynomial
${\mathcal L}[\ul\Xi]={\mathcal L}[\ul\Phi|\ul\Psi]$ (cf. \CMPref{2.8} for
the calculus of differential polynomials).
What we have is the action over bounded open space-time regions
$S_\Omega[\Xi]:=\int_\Omega dx{\mathcal L}[\Xi](x)\in\O(M)\sevR$, and we may
call
a $Z$-family of configurations $\Xi'=(\Phi'|\Psi')$ a {\em $Z$-family of
solutions} iff $(\chi S_\Omega)[\Xi'] =0$ for all vector fields $\chi$ on the
configuration smf which have their "target support" in the interior of
$\Omega$ (cf. \cite{[CMP4]} for precise definitions).

Now a $Z$-family of configurations $\Xi'$ is a family of solutions iff it
satisfies the arising field equations
\Beq
 \frac \delta{\delta\ul\Xi_i}{\mathcal L}[\Xi'] = 0
\Eeq
we call it a {\em $Z$-family of solutions.} Note that we take here
not functional derivatives, but variational ones; they are defined purely
algebraically.

Now, recalling that a $Z$-family $\Xi'$ is in essence the same as a morphism
$\check \Xi':Z\to M$, the smf of classical
solutions will be characterized by the following "universal property":

{\em A $Z$-family of configurations $\Xi'$ is a family of solutions iff the
corresponding morphism $\check\Xi':Z\to M$ factors through
the sub-smf $M\sol$.
}

It follows that $M\sol$ can be understood as the representing object for
the cofunctor
\Beq
 \Smfs\to \Sets,\quad
 Z\mapsto\{\text{$Z$-families of solutions}\}.
\Eeq

Roughly spoken, the superfunctions on $M\sol$ are just the
classical {\em on-shell observables} (i. e. equivalence classes
of observables with two observables being equivalent iff they
differ only by the field equations).

In the context of a model in Minkowski space, it is a natural idea that any
solution should be known by its Cauchy data, and that the "general"
solution, i. e. the universal solution family to be determined, should
be parametrized by all possible Cauchy data.

Thus, one introduces an {\em smf of Cauchy data} $M\Cau$ the underlying
manifold of which will be the manifold of ordinary Cauchy data for the
bosonic fields, and one constructs an smf morphism
\Beqn UnivSol
 \Xi\sol: M\Cau\to M
\Eeq
with the following property: Given a morphism $\zeta: Z\to M\Cau$,
i. e. effectively a $Z$-family of Cauchy data, the composite
$\Xi\sol\circ\zeta: Z\to M$ describes the unique $Z$-family of solutions
with these Cauchy data. Thus, $\Xi\sol$ is now the {\em universal family
of solutions} from which every other family of solutions
arises as pullback. The image of the morphism \eqref{UnivSol} will be just
the sub-smf $M\sol$ of solutions.

The successor papers \cite{[CMP3]}, \cite{[CMP4]} will implement this point
of view.

\subsubsection{Symmetries of the classical theory}
First, let us look at Euclidian models, where the global action
exists as a superfunction $S\in\O(M)\sevR$:

{\em The (super) Lie algebra of the infinitesimal symmetries of the
theory manifests itself as (super) Lie algebra of vector fields
on $M$ which leave $S$ invariant. These vector fields are tangential
to $M\sol$ and hence restrict to it.}

The discussion of \CMPref{3.13} yields a example for the appearance
of supersymmetry as an algebra of vector fields on a configuration
supermanifold (which, however, is so small that it allows
a globally defined action but no non-trivial solutions of the field
equations).

On the other hand, {\em on-shell (super-) symmetry manifests itself
as linear space of vector fields on $M$ which still leave $S$
invariant, but which only on $M\sol$ form a representation
of the super Lie algebra of supersymmetry.}

Unfortunately, in the Minkowski situation, the non-existence of a
globally defined action makes it difficult to give a general definition
of symmetries of the action; we will discuss that elsewhere.

Naively, one should expect that any infinitesimal (super-)
symmetry algebra integrates to an action $\alpha: G\times M\to M$ of
the corresponding connected, simply connected (super) Lie group
$G$ on $M$. For Yang-Mills gauge symmetry, this will work perfectly well;
on the other hand, for space-time symmetry groups,
there will arise the obstackle that we have only a real-analytic calculus
while $\alpha$ can only be expected to be smooth. A detailed
discussion will be given elsewhere.

\subsubsection{Outlook}\label{FQt:SymplSpek}
The classical, non-quantized version
of the canonical (anti-\nolinebreak[4]) commutation relations,
the canonical Poisson brackets
\Beqn CCR
 \left\{ \Xi_i(x),\Sigma_j(y)\right\}=\delta_{ij}\delta(x-y)
\Eeq
where
$\Sigma_j := \left(\partial L / \partial (\partial_0\ul\Xi_j)\right)[\Xi]$
are the canonical momenta, suggests the introduction of the two form
\Beq
 \omega\Cau = \sum_{i=1}^{N^\Phi+N^\Psi}
  \int_{\mathbb R^3} dx \delta\Xi_i(x) \delta\Sigma_i(x)
\Eeq
on the smf of Cauchy data $M\Cau$; here $\delta$ is the exterior derivative
for forms on the smf $M\Cau$  (cf. \cite{[IS2]} for a detailed theory),
and the product under the integral is the exterior product of one forms.

This equips the smf $M\Cau$ with a
symplectic structure, and \eqref{CCR} holds in "smeared" form:
\Beq
 \Bigl\{ \int dx f(x) \Xi_i(x), \int dx g(x) \Sigma_j(x)\Bigr\}
  =\int dx f(x)g(x)
\Eeq
for $f,g\in{\mathcal D}(\mathbb R^3)$ (all integrals over $\mathbb R^3$).
The necessity of the buffer functions
$f,g$ is connected with the fact that on an infinite-dimensional symplectic
(super-)manifold, the Poisson bracket is defined only on a subalgebra
of the (super-)function algebra. Thus, the non-smeared writing
\eqref{CCR} used in the textbooks of physicists is highly symbolic
since the r.h.s. is not a well-defined superfunction.

With the aid of the isomorphism $M\Cau \too {\Xi\sol} M\sol$, one can
carry over $\omega\Cau$ to a symplectic structure $\omega\sol$ on
$M\sol$; we will show elsewhere that this is Lorentz invariant, and hence an
intrinsic structure (cf. also \cite{[Seg]} for an alternative construction.
This paper constructs heuristically a pseudo-K\"ahler structure on
$M\sol$; however, the well-definedness of the latter is not clear).

This symplectic structure  makes it possible to rewrite the field
equations in Hamiltonian form. In view of this, it is natural to call
the smf $M\sol$ also the {\em covariant phase space} of the theory
(cf. \cite[17.1.2]{[Hen/Tei]}, \cite{[Crn]}).

The symplectic smf $M\Cau$ might be the starting point for a geometric
quantization. Of course, it is a rather tricky question what the
infinite-dimensional substitute for the symplectic volume needed for
integration should be; we guess that it is some improved variant of
Berezin's functional integral (cf. \cite{[Ber2ndQ]}). Note, however,
that although the integration domain $M\sol$ is isomorphic to the linear
supermanifold $M\Cau$, this isomorphism is for a model with interaction
highly Lorentz-non-invariant, and Berezin's functional integral makes
use of that linear structure.

Also, in the interacting case, the usual problems of quantum field theory, in
particular renormalization, will have to show up on this way, too, and
the chances to construct a Wightman theory are almost vanishing. Nevertheless,
it might be possible to catch some features of the physicist's computational
methods (in particular, Feynman diagrams),
overcoming the present mathematician's attitude of contempt and disgust to
these methods, and giving them a mathematical description
of Bourbakistic rigour.

What certainly can be done is a mathematical derivation of the rules which lead
to the tree approximation $S_{\opn tree }$ of the scattering operator.
$S_{\opn tree }$ should be at least a well-defined power series; of course,
the wishful result is that in a theory without bounded states,
$S_{\opn tree }$ is defined as an automorphism
of the solution smf $M^{\opn free }$ of the free theory.

\subsection{Example: Fermions on a lattice}\label{Lattice}
As an interesting example for the description of
configuration spaces as supermanifolds, we re-describe the standard
framework of Euclidian Yang-Mills-Higgs-Dirac theory on a finite lattice
(cf. \cite{[Seiler]}) in supergeometric terms. Here the finite-dimensional
smf calculus is still sufficient.

Let $\Lambda\subset\mathbb Z^d$ be a finite lattice.
Let $L^*:=\{(x,y)\in\Lambda^2:\ \br\|x-y\|=1\}$ the set of its
links, and let $L:=\{(x,y)\in L^*:\ \forall i\ \ y_i\ge x_i\}$.

Let $G$ be a compact Lie group; we recall that $G$ has a canonical
real-analytic structure.

A {\em configuration of the gauge field} is a map
$g: L^*\to G$,\ \ $(x,y)\mapsto g_{xy}$ with $g_{xy}g_{yx}=1$; thus, the
configuration space is the real-analytic manifold $G^L$.
The {\em Wilson-Polyakov action} is the real-analytic function
\Beq
 S_{\opn YMW }: G^L\to \mathbb R,\quad
 g\mapsto - \frac 12 \sum \chi(g_{vw}g_{wx}g_{xy}g_{yu})
\Eeq
where the sum runs over all plaquettes of $\Lambda$, i. e. all tuples
$(x,y,u,v)\in G^4$ for which
$(x,y),\allowbreak (y,u)\in L$, $(u,v),\allowbreak (v,w)\allowbreak \in L^*$;
tuples which differ only by cyclic permutations are identified.
Also, $\chi: G\to \mathbb R$ is some character of $G$ belonging to a
locally exact representation. (We ignore all wave function renormalization
and coupling constants.)

For the {\em Higgs field}, we need a finitedimensional real Hilbert space
$\cV_H$ and an orthogonal representation
$U\Higgs: G\to\Ort(\cV\Higgs)$. A {\em configuration of the Higgs field}
is a map $\phi: \Lambda\to \cV\Higgs$; thus, the configuration space for the
Higgs field is the vector space ${\cV\Higgs}^\Lambda$.
The Higgs action is the real-analytic function
\Beq
 M_{\opn bos }= G^L\times {\cV\Higgs}^\Lambda \too {S\Higgs} \mathbb R,\quad
 (g,\phi)\mapsto -\frac 12
 \sum_{(x,y)\in L} (\phi(x), U\Higgs(g_{xy})\phi(y)) + \sum_{x\in\Lambda}
 V(\br|\phi(x)|)
\Eeq
where $V$ is a fixed polynomial of even degree $\ge4$ with positive highest
coefficient.

Turning to the Dirac field, we assume to be given a Clifford module $\cV\Spi$
("spinor space" in \cite{[Seiler]}) over $\opn Cliff (\mathbb R^{d+1})$; that
is,
$\cV\Spi$ is a finitedimensional complex Hilbert space together with
selfadjoint operators $\gamma_i\in\opn End _{\mathbb C}(\cV\Spi)$\ \
($i=0,\dots,d$) such that $\gamma_i\gamma_j+\gamma_j\gamma_i=2\delta_{ij}$.

Also, we need a finitedimensional real Hilbert space
$\cV\Gauge$ ("gauge" or "colour" space) and an orthogonal representation
$U\Gauge: G\to\Ort(\cV\Gauge)$. Let
$\cV\Ferm:=\cV\Spi\otimes_{\mathbb R} \cV\Gauge$, which
is a complex Hilbert space, and consider it purely odd. Let $(e_\alpha)$,
$(f_a)$ be orthonormal bases of $\cV\Spi$ and $\cV\Gauge$, respectively; thus,
the elements
\Beq
 \Psi_{\alpha a}:=e_\alpha\otimes f_a
\Eeq
form an orthonormal basis of $\cV\Ferm$.
Now take one copy $\cV\Ferm(x)$ of $\cV\Ferm$ for each lattice site
$x\in\Lambda$, with basis $(\Psi_{\alpha a}(x))$, and let $\cj{\cV\Ferm(x)}$
be the "exterior" conjugate of $\cV\Ferm(x)$, i.e. the Hilbert space
consisting of all elements $\cj v$ with $v\in\cV\Ferm(x)$; Hilbert space
structure and $G$-action  are fixed by requiring that the map
$\cV\Ferm(x)\to\cj{\cV\Ferm(x)}$, \ \ $v\mapsto \cj v$, be
antilinear, norm-preserving, and $G$-invariant.

Setting
\Beq
 \cV:= \bigoplus_{x\in\Lambda} \cV\Ferm(x),
 \quad\cj{\cV}:= \bigoplus_{x\in\Lambda} \cj{\cV\Ferm(x)}
\Eeq
(of course, the "$\otimes$" of \cite{[Seiler]} is a misprint),
the space $\cV\oplus \cj{\cV}$ has a natural hermitian structure
(conjugation acts as the notation suggests). Let
$\cV_r:=\{v+\cj v;\ v\in \cV\}$ be its real part; thus, $\cV$ identifies
with the complexification of $\cV_r$. Now the Grassmann algebra
\Beq
 {\mathcal G}_\Lambda :=
 \bigwedge(\cV\oplus \cj{\cV}),
\Eeq
is nothing but the algebra of superfunctions on the hermitian supermanifold
$M_{\opn ferm }:=\L(\cV_r)$:
\Beq
 {\mathcal G}_\Lambda = \O(M_{\opn ferm })
\Eeq
(recall that $\cV_r$, being a real Hilbert space, identifies canonically
with its dual), thus, $M_{\opn ferm }$ is the configuration supermanifold
for the fermionic field strengthes. The field strengthes now appear as
odd superfunctions:
\Beq
 \Psi_{\alpha a}(x),\cj{\Psi_{\alpha a}(x)}\in\O(M_{\opn ferm })\sodd;
\Eeq
in the language of \cite{[HERM]}, the $\Psi_{\alpha a}(x)$ form a
chiral coordinate system on $M_{\opn ferm }$.

The configuration space for the whole system is the hermitian
supermanifold
\Beq
   M = M_{\opn bos }\times M_{\opn ferm }
   = G^L\times {\cV\Higgs}^\Lambda \times \L(\cV_r).
\Eeq
The algebra of global superfunctions is the Grassmann algebra generated by
$\Psi_{\alpha a}(x), \cj{\Psi_{\alpha a}}(x)$ with the
coefficients being real-analytic functions of the $g_{xy}$, $\phi(x)$.

(Actually, \cite{[Seiler]} uses instead of this the Grassmann algebra
${\mathcal A}_\Lambda$ with the same generators but with the coefficients
being continuous bounded functions of the $g_{xy}$, $\phi(x)$; but that
difference does not really matter.)

Turning to the fermionic action, fix parameters $r\in(0,1]$,
$\theta\in[0,\pi/2)$; for their meaning, we refer to the literature.
For $(x,y)\in L$ set
\Beq
   \Gamma^{xy}_{\alpha\beta} := r\exp(\i\theta\gamma_d) +
   \sum_{j=0}^d (x_j-y_j)(\gamma_j)_{\alpha\beta};
\Eeq
since $x,y$ are neighbours, only one term of the sum is non-zero.
Also, observe that $g\mapsto U\Gauge(g_{xy})_{ab}$ is a real-analytic function
on $G^L$. Therefore we can form the superfunction
\Bml
 S\Ferm := \frac 12 \sum_{(x,y)\in L} \sum_{a,b,\alpha,\beta}
  \cj{\Psi_{\alpha a}(x)}
  \Gamma^{xy}_{\alpha\beta} U\Gauge(g_{xy})_{ab}
  \Psi_{\beta b}(y)
  \\
  + \frac 12 \sum_{x\in\Lambda} \sum_{a,\alpha,\beta}
  \cj{\Psi_{\alpha a}(x)}
  \Bigl(M - rd\exp(\i\theta\gamma_d)_{\alpha\beta}\Bigr)
  \Psi_{\beta a}(x),
\Eml
which is the action for the fermion field. Here $M$ is the mass of the Dirac
field.

The total action is now a superfunction on $M$:
\Beq
 S_{\opn tot }:= S_{\opn YMW } + S\Higgs + S\Ferm \in\O(M)\seven.
\Eeq
We look at the action of the group of gauge transformations
$G^\Lambda$, which is the Lie group of all maps $h: \Lambda\to G$: By
\Beq
 (h\cdot g)_{xy}:= h_xg_{xy}{h_y}^{-1},\quad
 (h\cdot\phi)(x):=  U\Higgs(h_x)\phi(x),
\Eeq
we get an action
\Beqn GGonGV
 G^\Lambda \times M_{\opn bos }\to M_{\opn bos }
\Eeq
of the group of gauge transformations on the configuration space for
the gauge and Higgs fields. On the other hand, we have a representation
of $U\Ferm: G^\Lambda\to\Ort(\cV_r)$ by restricting the natural unitary
action of $G^\Lambda$ on $\cV$ to the real
part (this works due to the reality of the representation $U\Gauge$).
We get a representation (cf. \cite[5.18-20]{[IS2]} for some basic notions)
\Beqn GGonL
 a: G^\Lambda \times M_{\opn ferm }\to M_{\opn ferm },\quad
 a^*(\Psi_{\alpha a}(x)) = U\Gauge(h_x)\Psi_{\alpha a}(x)
\Eeq
of $G^\Lambda$ on the linear supermanifold $M_{\opn ferm }$.
\eqref{GGonGV}, \eqref{GGonL} together yield an action
\Beq
 a: G^\Lambda \times M \to M
\Eeq
of the group of gauge transformations on the configuration supermanifold,
and $S$ is now an invariant function, i.e. $a^*(S) = {\pr_2}^*(S)$.

Let $\O_b(M_{\opn bos })$ be the subspace of all $f\in\O(M_{\opn bos })$
which grow only polynomially in Higgs direction, i. e. there exist $C,p>0$
such that
\Beq
   \br|f(g,\phi)(x)| \le C(1+\br|\phi(x)|^p)
\Eeq
for $x\in\Lambda$, and let
\Beq
 \O_b(M) := \O_b(M_{\opn bos })\otimes \O(M_{\opn ferm })\seq \O(M).
\Eeq
Now we can interpret the {\em mean value} as a Berezin integral:
For $P\in\O_b(M)$, we may form
\Beqn MeanVal
 \br<P>_\Lambda := \int dg \prod_{k,x} d\phi_k(x)
 \int \prod_{\alpha,a,x} d\Psi_{\alpha a}(x)d\cj{\Psi_{\alpha a}(x)}
 \ \ P\exp(-S_{\opn tot }/\hbar).
\Eeq
Here the inner integral is Berezin integration along the fibres of
the super vector bundle
$M= M_{\opn bos }\times\L(\cV_r) \too\pr M_{\opn bos }$,
producing an ordinary function on $M_{\opn bos }$.

The $\phi_k(x)$ ($k=1,\dots,\dim \cV\Higgs$) are orthonormal coodinates on
the $x$-component of ${\cV\Higgs}^\Lambda$; thanks to the exponential factor
and our growth condition, the integral over them is finite.
Finally, $dg$ is the normalized Haar measure on $G^L$.
The sign ambiguity arising from the missing order of the $\phi_k(x)$ is
resolved by fixing $\br<1>_\Lambda>0$.

Now we suppose the existence of {\em time reflection} as a fixpoint-free
involutive map $r:\Lambda\to\Lambda$ which respects the link structure,
$(r\times r)(L^*)\seq L^*$. By permuting the factors, $r$ yields an involution
$r: M_{\opn bos }\to M_{\opn bos }$. Also, we get a morphism
\Beq
 r: M_{\opn ferm }\to M_{\opn ferm },\quad
 r^*(\Psi_{\alpha a}(x))=
  \sum_\beta \Psi_{\beta a}(rx)(\gamma_0)_{\alpha\beta}.
\Eeq
Altogether, we get an involutive morphism $r: M\to M$ and a new hermitian law
\Beq
 \theta(P):= \cj{r^*(P)}.
\Eeq
(Actually, the notation is unlucky in view of \ref{HermStuff}, because
it hides the skew-linearity.) We also need a decomposition of the lattice
\Beq
 \Lambda=\Lambda_+ \cup \Lambda_-, \quad \Lambda_+ \cap \Lambda_-=\emptyset
\Eeq
such that $r(\Lambda_\pm)\seq\Lambda_\mp$ (cf. \cite{[Seiler]} for the
background). We get a projection morphism $M=M_\Lambda \to M_{\Lambda_+}$,
and hence an embedding $\O_b(M_{\Lambda_+})\seq \O_b(M)$.
{\em Osterwalder-Schrader positivity} now states that
\Beq
    \br<P\theta(P)> \ge0
\Eeq
for all $p\in\O_b(M_{\Lambda_+})$. Also, the scalar product
\Beq
 \O_b(M_{\Lambda_+})\times\O_b(M_{\Lambda_+})\to \mathbb C,\quad
 (P,Q)\mapsto\br<P\theta(Q)>,
\Eeq
satisfies
\Beqn ScalPrd
  \br<Q\theta(P)> = \cj{\br<P\theta(Q)>};
\Eeq
hence, it equips $\O_b(M_{\Lambda_+})/\{P: \br<P\theta(P)>=0\}$ with the
structure of a pre-Hilbert space the completion ${\mathcal H}$ of which is the
{\em Euclidian state space}. (Thus, $\O_b(M_{\Lambda_+})$ plays here the same
role as the space of polarized sections in geometric quantization.)

\Brm
(1) Note that working with a conventional, non-hermitian calculus would bring
trouble here since \eqref{ScalPrd} would acquire an additional factor
$(-1)^{\br|Q|\br|P|}$, and hence $\br<P\theta(P)>$ would be for odd $P$
not positive but imaginary.

(2) It would be interesting to know something about the "supermanifold of
classical solutions" of the action $S$. That is, we may form the ideal sheaf
${\mathcal J}(\cdot)\seq\O_M(\cdot)$ generated by the "field equations"
\Beq
 \frac \partial{\partial g_{xy}^i} S,\ \
 \frac \partial{\partial \phi_k(x)}S,\ \
 \frac \partial{\partial \Psi_{\alpha a}(x)}S,\ \
 \frac \partial{\partial\cj{\Psi_{\alpha a}(x)}}S\ \in \O(M),
\Eeq
(here $g_{xy}^i$ are local coodinates on the $(x,y)$-component of $G^L$)
and take the factor space $M\sol := (\opn supp \O/{\mathcal J},\ \ \O/{\mathcal
J})$.
This is at least a real superanalytic space which, however, might have
singularities. Note that, in contrast to the non-super situation, where
the singular locus of an analytic space has codimension $\ge1$,
it may on a superanalytic space be the whole space; it would be nice to show
that this does not happen here.

Also, one should prove that in the classical limit $\hbar\to0$,
the integral \eqref{MeanVal} becomes asymptotically equal
to an integral over $M\sol$ (or its non-singular part). Indeed, in the bosonic
sector, the exponential factor makes the measure accumulate on the
subspace $\widetilde{M\sol}$ of the minima of the action; however, in the
fermionic sector, the picture is less clear.
\Erm

\section{Infinite-dimensional supermanifolds}\label{InfDSmfs}
An attentive reader of \cite{[CMP1]} will have noted that most of its material
(with the exception of local functionals, differential polynomials a.s.o.)
does not really depend on the fact that $E$ is a function space. In
fact, the theory can be developed in an abstract context, and one really
should do so in order to get conceptual clarity which may be useful if
a concrete situation does not fit into our framework. In that way, we
will also establish the connection with usual, finite-dimensional
supergeometry a la Berezin.

Some work has been done on infinite-dimensional supergeometry and its
application onto classical fields in quantum field theory.
Apart from the implicite appearance of infinite-dimensional supermanifolds
in \cite{[Ber2ndQ]} (cf. \cite{[Rapallo]}), the first work on the mathematical
side is Molotkov \cite{[Mol 84]}; however, his approach is not well suited
for physical purposes. Cf. below for a discussion.

\cite{[Abramov 86]} uses an ad-hoc definition of smooth Banach smfs in
order to describe mathematically the fermionic Faddeev-Popov ghost fields
used by physicists for quantization of the Yang-Mills field.

In \cite{[Kostant-Sternberg 87]}, infinite-dimensional supergeometry makes
an implicite appearance in a more general approach to the quantization
of systems with first-class constraints (cf. also the comment in
\cite{[Ewen-Schaller-Schwarz 1991]}).

\cite{[Schaller-Schwarz 1990]} and \cite{[Ewen-Schaller-Schwarz 1991]}
use, on a physical level of rigour, ad-hoc generalizations of the
Berezin-Kostant supergeometry framework for applying geometric
quantization in field theory, in particular to fermionic fields.

Finally, the present author constructed in \cite{[ISA]},
\cite{[IS1]}, \cite{[IS2]},
the predecessors of the present paper, a rather general theory of complex-
and real-analytic supermanifolds modelled over locally convex spaces.

Here, we will give an alternative description of this theory, using a
traditional treatment via ringed spaces and charts. Also, we
will treat only real-analytic supermanifolds with complete
model spaces.

We will give only a short account on the abstract variant since the
details should be clear from the material of \cite{[CMP1]}.

\subsection{Formal power series}\label{FPSer}
Let $E$, $F$ be complete \ztwo-lcs, and define the space
$\P^{k|l}(E;F)$ of {\em $F$-valued $k|l$-forms on $E$} as the space of
all $(k+l)$-multilinear continuous maps
\Beq
 u_{(k|l)}:\prod^kE\seven\times\prod^lE\sodd\to F_{\mathbb C}
\Eeq
which satisfy the symmetry requirement
\Beq
 u_{(k|l)}(e_{\sigma(1)},\dots,e_{\sigma(k)},e'_{\pi(1)},\dots,e'_{\pi(l)})
 =\opn sign (\pi)u_{(k|l)}(e_1,\dots,e_k,e'_1,\dots,e'_l)
\Eeq
for all permutations $\sigma,\ \pi$. $\P^{k|l}(E;F)$ is a
\ztwo-graded vector space; note that we do not distinguish a
topology on it.

The space of {\em F-valued formal power series} on $E$ is defined by
\Beq
 \P_f(E;F):=\prod_{k,l\ge0}\P^{k|l}(E;F);
\Eeq
thus, its elements are formal sums $u=\sum_{k,l\ge0}u_{(k|l)}$ where
$u_{(k|l)}\in\P^{k|l}(E;F)$. The rule
\Beq
 \cj u_{(k|l)}(e_1,\dots,e_k,e'_1,\dots,e'_l):=
 \cj{u_{(k|l)}(e_1,\dots,e_k,e'_l,e'_{l-1},\dots,f_1)}
\Eeq
\Beq
 =(-1)^{\binom l2}\cj{u_{(k|l)}(e_1,\dots,e_k,e'_1,\dots,e'_l)}
\Eeq
turns $\P_f(E;F)$ into a hermitian vector space. The {\em product}
\Beq
 \P^{k|l}(E;F)\times\P^{k'|l'}(E;F')\to\P^{k+k'|l+l'}(E;F\widehat\otimes F')
\Eeq
is defined by
\Bal
 &(u\otimes v)_{(k+k'|l+l')}(e_1,\dots,e_{k+k'},e'_1,\dots,e'_{l+l'})=
\\
 &\quad=\sum(\pm) \binom{k+k'}k^{-1} \binom{l+l'}l ^{-1}
 u_{(k|l)}(e_{p_1},\dots,e_{p_k},e'_{q_1},\dots,e'_{q_l})\otimes
 v_{(k'|l')}(e_{p'_1},\dots,e_{p'_k},e'_{q'_1},\dots,e'_{q'_l}).
\Eal
(the sign "$\otimes$" on the l.h.s. is somewhat abusive).
Here the sum runs over all $\binom{k+k'}k \binom{l+l'}l$ partitions
\Beq
 \{1,\dots,k+k'\}=\{p_1,\dots,p_k\}\sqcup\{p'_1,\dots,p'_{k'}\},\quad
 p_1\le\dots\le p_k,\quad
 p'_1\le\dots\le p'_{k'}
\Eeq
\Beq
 \{1,\dots,l+l'\}=\{q_1,\dots,q_l\}\sqcup\{q'_1,\dots,q'_{l'}\},\quad
 q_1\le\dots\le q_l,\quad
 q'_1\le\dots\le q'_{l'}
\Eeq
and $(\pm)$ is given by the sign rule:
\Beq
 (\pm):=(-1)^{|v|(|q_1|+\dots+|q_l|)}\opn sign (\pi)
\Eeq
where $\pi$ is the permutation $(q_1,\dots,q_l$, $q'_1,\dots,q'_{l'})$
of $\{1,\dots,l+l'\}$ (cf. \cite[Prop. 2.3.1]{[IS1]}).

The product turns $\P_f(E;\mathbb R)$ into a \ztwo-commutative
hermitian algebra and each $\P_f(E;F)$ into a hermitian module over
that algebra. In both situations, we will usually write simply $uv$
instead of $u\otimes v$.

\Brm
An $F$-valued formal power series in the sense of \CMPref{2.3},
\Beq
 K[\Phi|\Psi]=\sum_{k,l\ge0} \frac 1{k!l!} \sum_{I|J} \int dXdY K^{I|J}(X|Y)
 \prod^k_{m=1} \Phi_{i_m}(x_m) \cdot \prod^l_{n=1} \Psi_{j_n}(y_n)
 \ \in{\mathcal P}(F),
\Eeq
defines an $F$-valued formal power series $K\in\P(D;F)$ in the sense above,
$K=\sum_{k,l} K_{(k|l)}$ with
\Bmln TklMap
 K_{(k|l)}: \prod^k D\seven\times \prod^l D\sodd\to F_{\mathbb C},\qquad
 (\phi^1,\dots,\phi^k,\psi^1,\dots,\psi_l) \mapsto
 \\
 \frac 1{k!l!} (-1)^{l(l-1)/2} \int dXdY \sum_{I|J} K^{I|J}(X|Y)
 \prod_{m=1}^k\phi^m_{i_m}(x_m)\cdot\prod_{n=1}^l\Pi\psi^n_{j_n}(y_n)
\Eml
where $D={\mathcal D}(\mathbb R^d)\otimes V$, and $V=V\seven\oplus V\sodd$ is
the
field target space. (The apparently strange parity shift $\Pi$ was motivated
by the wish to have $\Psi$ as an odd symbol.)

Moreover, if $E$ is an admissible function space in the sense of
\CMPref{3.1}, i.~e. $E$ is a \ztwo-graded complete locally convex space
with continuous inclusions $D\seq E\seq{\mathcal D}'(\mathbb R^d)\otimes V$,
then
we have $K\in\P_f(E;F)$ iff for all $k,l\ge0$,  \eqref{TklMap}
extends to a continuous map
$\bot^k E\seven\otimes \bot^l E\sodd\to F_{\mathbb C}$. In that way, we get
a natural identification between $\P_f(E;F)$ as defined in \CMPref{3.1} and
the $\P_f(E;F)$ defined here.

(In \CMPref{3.1}, we had also assigned to $K_{(k|l)}[\Phi|\Psi]$ the linear
map
\Beq
 \bot^k D\seven\otimes \bot^l\Pi D\sodd\to F_{\mathbb C},\qquad
 \bot_{m=1}^k \phi^m \otimes \bot_{n=1}^l \Pi\psi^n \mapsto
 (-1)^{l(l-1)/2}\cdot(\text{second line of \eqref{TklMap}}).
\Eeq
Note, however, that the parity of this map differs from that of
$K[\Phi|\Psi]$ by the parity of $l$.)
\Erm

\subsection{Analytic power series}
Let be given continuous seminorms $p\in\CS(E)$, $q\in\CS(F)$.
We say that $u\in\P_f(E;F)$ {\em satisfies the $(q,p)$-estimate} iff
we have for all $k,l$ with $k+l>0$ and all $e_1,\dots,e_k\in E\seven$,
$e'_1,\dots,e'_l\in E\sodd$ the estimate
\Beq
 q(u_{(k|l)}(e_1,\dots,e_k,e'_1,\dots,e'_l))
 \le p(e_1)\cdots p(e_k)p(e'_1)\cdots p(e'_l)
\Eeq
(we extend every $q\in\CS(F)$ onto $F_{\mathbb C}$ by
$q(f+\opn i f'):=q(f)+q(f')$). We call $u$ {\em analytic} iff for each
$q\in\CS(F)$ there exists a $p\in\CS(E)$ such that $u$ satisfies the
$(q,p)$-estimate.

Now every $k|l$-form $u_{(k|l)}\in\P^{k|l}(E;F)$ is analytic, due
to its continuity property, and analyticity of a formal power
series is just a joint-continuity requirement onto its coefficients
$u_{(k|l)}$.

The analytic power series form a hermitian subspace $\P(E;F)$
of $\P_f(E;F)$. Moreover, tensor product of analytic power series
and composition with linear maps in the target space produce
analytic power series again.

For $e"\in E$, the {\em directional derivative} $\partial_{e"}$
is defined by
\Beq
 (\partial_{e"}u)_{(k|l)}(e_1,\dots,e_k,e'_1,\dots,e'_l)
 :=\begin{cases}
  (k+1)u_{(k+1|l)}(e",e_1,\dots,e_k,e'_1,\dots,e'_l)
   & \text{for $|e"|=0$}\\
  (-1)^{|u|}(l+1)u_{(k|l+1)}(e_1,\dots,e_k,e"_1,e'_1,\dots,e'_l)
   & \text{for $|e"|=1$.}
   \end{cases}
\Eeq
$\partial_{e"}$ maps both $\P_f(E;F)$ and $\P(E;F)$ into themselves,
and it acts as derivation on products:
\Beq
  \partial_e(u\otimes v) = \partial_eu\otimes v
  + (-1)^{\br|e|\br|u|}u\otimes \partial_ev.
\Eeq
The abstract analogon of the functional derivative of $K$ is the
linear map
\Beq
 E\to\P_f(E;F),\qquad  e\mapsto\partial_eK.
\Eeq

\medskip

Suppose that $F$ is a \ztwo-graded Banach space, and fix
$p\in\CS(E)$. Set
\Beq
 \|u_{(k|l)}\|_p:=\opn inf  \{c>0:\ \
 \text{$u_{(k|l)}$ satisfies a ($c^{-1}\|\cdot\|,p)$-estimate} \}
\Eeq
for $k+l>0$, and define $\|\cdot\|$ on $0|0$-forms to be the norm
in $F_{\mathbb C}$. Then
\Beq
 \P(E,p;F)=\{u\in\P(E;F):\ \ \|u\|_p:=\sum_{k,l\ge0}\|u_{(k|l)}\|<\infty\}
\Eeq
is a Banach space. Moreover, for any $E$ we have
\Beq
 \P(E;F)=\bigcup_{p\in\CS(E)}\P(E,p;F).
\Eeq
Now if $F'$ is another \ztwo-graded Banach space then 
\Beq
 \|u\otimes v\|_p \le\|u\|_p \|v\|_p
\Eeq
for $u\in\P(E;F)$, $v\in\P(E;F')$. In particular, $\P(E,p;\mathbb R)$
is a Banach algebra.

Note that for every $p\in\CS(E)$, $c>1$, the directional derivative
$\partial_e$ maps $\P(E,p;F)\to\P(E,cp;F)$ for any $e\in E$.

\Brm
(1) A $k|l$-form $u_{(k|l)}$ lies in $\P(E,p;F)$
iff it factors through $\hat E_p$; in that case, $\|u_{(k|l)}\|$
is just its supremum on the $k+l$-fold power of the unit ball of this space.

(2) Remark 3.2.2 of \cite{[CMP1]} carries over, linking the approach here
with \cite{[IS1]}: Fixing $u\in\P_f(E;F)$ and $k,l\ge0$ we get a
continuous map
\[
 \S^k E_{\even,\mathbb C}\cdot \S^l E_{\odd,\mathbb C}\to F_{\mathbb C},\qquad
 e_1\cdots e_k e'_1\cdots e'_l
 \mapsto
 k!l! u(e_1,\dots,e_k,e'_1,\dots,e'_l)
\]
(using notations of \cite{[IS1]}; the topology on the l. h. s. is induced
from the embedding into $\S E_{\mathbb C}$).
Using Remark 2.1.(2) of \cite{[IS1]} we get a bijection
\[
 \P_f(E;F)\to \prod_{k\ge0} {\mathcal L}(\S^k E_{\mathbb C},F_{\mathbb C})
\]
(symmetric algebra in the super sense). The r. h. s. is somewhat
bigger than ${\mathcal L}(\S E_{\mathbb C},F_{\mathbb C})={\opn P }(E;F)$,
due to the absence of growth conditions. Having identified both sides,
one shows for $u\in\P_f(E;F)$ the estimates
\[
 \Vert u\Vert\dn{U_{p/2}} \le \Vert u\Vert_p \le \Vert u\Vert\dn{U_{2p}}
\]
(cf. \cite[2.5]{[IS1]} for the notations
$\Vert\cdot\Vert\dn U,\ {\opn P }(E,U;F)$), and hence
\[
 {\opn P }(E,U_{p/2};F) \seq \P(E,p;F)\seq{\opn P }(E,U_{2p};F),
 \qquad \P(E;F) = {\opn P }(E;F).
\]
*
\Erm

\subsection{Insertions}\label{Insts}
On the level of formal power series,
we will define $u[v]\in\P_f(E';F)$ with the data
\Beq
 u\in\P_f(E;F),\qquad
 v\in\P_f(E';E)\seven,\qquad
 v_{(0|0)}=0.
\Eeq
We split $v=v_{(\even)}+v_{(\odd)}$ with $v_{(\mathbf i)}
 \in\P_f(E';E_{\mathbf i})\seven$ $(\mathbf i=\mathbf 0,\mathbf 1)$, and
we set
\Beqn InsertedSer
 u[v]:=\sum_{k,l\ge0}\Bigl<u_{(k|l)},
 \bigotimes^k v_{(\mathbf 0)}\otimes\bigotimes^lv_{(\mathbf 1)}\Bigr>.
\Eeq
Here $\bigotimes^kv_{(\mathbf 0)}\otimes\bigotimes^lv_{(\mathbf 1)}
\in\P_f(E';\bigotimes^kE\seven\otimes\bigotimes^l E\sodd)$ is the
product, and $u_{(k|l)}$ is viewed as continuous
linear map $\bigotimes^kE\seven\otimes\bigotimes^l E\sodd\to F_{\mathbb C}$,
\Beq
 \left<u_{(k|l)},
 e_1\otimes\cdots\otimes e_k\otimes e'_1\otimes\cdots\otimes e'_l\right>
 :=u_{(k|l)}(e_1,\dots,e_k,e'_1,\dots,e'_l).
\Eeq
Thus, each term of \eqref{InsertedSer} makes sense as element of $\P_f(E';F)$,
and by the same arguments as in \CMPref{3.3}, \eqref{InsertedSer}
is a well-defined formal power series.

If we want to lift the condition $v_{(0|0)}=0$ we have to introduce
more hypotheses. Suppose that

(1) $v\in\P(E';E)\seven$, \quad $u\in\P(E,p;F)$ where $F$ is a Banach
space and $p\in\CS(E)$, and

(2) there exists some $q\in\CS(E')$ with
\Beqn IpVInPEQ
 i_p v\in\P(E',q;\hat E_p),
\Eeq
\Beqn AbsTermLe1
 \|i_pv\|_q<1.
\Eeq
Defining $u_{(k|l)}[v]\in\P(E';F)$ as above, one has
$\|u_{(k|l)}[v]\|\le\|u_k[v]\|_p\cdot \|i_pv_{(\mathbf 0)}\|_q^k \cdot
 \|i_pv_{(\mathbf 1)}\|_q^l$, and therefore the series
\eqref{InsertedSer} converges in $\P(E',q;F)$.

Thus, for varying $u$, we get a linear map of Banach spaces
\Beq
 \P(E,p;F)\to\P(E',q;F),\quad u\mapsto u[v],
\Eeq
of norm $\le1/(1-\|i_pv\|_q)$; for $F=\mathbb R$, this is a homomorphism
of Banach algebras.

More generally, let again the data (1) be given, and replace (2) by
the weaker condition
\Beq
 p(v_{(0|0)})<1.
\Eeq
Then $u[v]\in\P(E';F)$ still makes sense: We can choose $q\in\CS(E')$
with \eqref{IpVInPEQ}. Now, replacing $q$ by $cq$ with a suitable $c>1$,
we can achieve \eqref{AbsTermLe1}.

Using these results it is easy to show that the insertion $u[v]$ where
$u,v$ are analytic and $v_{(0|0)}=0$ is analytic again.

\Brm
The connection of the insertion mechanism with
\cite{[IS1]} is the following: If $u,v$ are formal power series and
$v_{(0|0)}=0$ then $u[v]$ here is nothing but $u\circ\exp(v)$ of
\cite[2.3]{[IS1]}. More generally, if $u,\ v$ are analytic and
$\phi:=v_{(0|0)}\ne0$ then $u[v]=(\t_\phi u)\circ\exp(v-\phi)$ where
$\exp(v-\phi)$ is the cohomomorphism
$\opn S E'_{\mathbb C}\to\opn S E_{\mathbb C}$ determined by $v-\phi$.
\Erm

Whenever insertion is defined, it is associative, i. e. $u[v[w]]$
makes unambiguous sense if it is defined. (A direct proof is
sufficiently tedious. Cf. the approach of \cite{[IS1]} which saves that
work.)

The power series $x=x_E\in\P(E;E)$ defined by
\Beq
 x_{(1|0)}(e\seven)=e\seven,\qquad
 x_{(0|1)}(e\sodd)=e\sodd
\Eeq
$(e_{\mathbf i}\in E_{\mathbf i})$, $x_{(k|l)}=0$ for all other $k,l$, is the
unit
element under composition: for $u\in\P(E;F)$,
\Beq
 u[x_E]=x_F[u]=u.
\Eeq
$x$ is the abstract analogon to the $\Xi$ considered
in \CMPref{3.3}. Sometimes, we call $x$ the {\em standard coordinate},
and we will write power series in the form $u=u[x]$.

Suppose again that $F$ is a Banach space, and let
\Beq
 u\in\P(E,p;F),\qquad
 e\in E\seven,\qquad
 c:=1-p(e)>0.
\Eeq
We define the {\em translation} $\t_eu\in\P(E,cp;F)\seq\P(E;F)$ by
\Beq
 \t_eu[x]:=u[x+e].
\Eeq
Here $e$ is viewed as constant power series $e\in\P(E;E\seven)$. We have
the {\em Taylor formula}
\Beq
 \t_eu =\sum_{k\ge0}\frac 1{k!}\partial_e^ku;
\Eeq
the sum absolutely converges in $\P(E,cp;F)$. Thus
\Beq
 (\t_eu)_{(k|l)}(e_1,\dots,e_k,e'_1,\dots,e'_l)
 =\sum_{k'\ge0}\binom{k+k'}k u_{(k+k'|l)}(
 \underbrace
 {e,\dots,e}_{\text{$k'$ times}},e_1,\dots,e_k,e'_1,\dots,e'_l).
\Eeq
Sometimes, we will write $u\circ v$ instead of $u[v]$, and call this the
{\em composition of $u$ with $v$.}

\subsection{Superfunctions} \label{SFcts}
Let $E, F$ be complete \ztwo-lcs and $U\seq E\seven$ open. The material of
\CMPref{3.5} carries over verbally. Thus, an {\em $F$-valued (real-analytic)
superfunction on $U$} is a map
\Beqn SFctCand
 u:U\to\P(E;F),\qquad e\mapsto u_e,
\Eeq
which satisfies the following condition: whenever $u_e$ satisfies
a $(q,p)$-estimate we have for all $e'\in U$ with $p(e'-e)<1$
\Beq
 i_qu_{e'}=\t_{e'-e}i_qu_e.
\Eeq
We call $u_e$ the {\em Taylor expansion} of $u$ at $e$.

We get a sheaf $\O^F(\cdot)$ of hermitian vector spaces on $E$,
and the product globalizes to a bilinear map
\Beq
 \O^F(U)\times\O^G(U)\to\O^{F\otimes G}(U), (u,v)\mapsto u\otimes v.
\Eeq
In particular, the sheaf $\O(\cdot):=\O^{\mathbb R}(\cdot)$
of {\em scalar superfunctions} is a sheaf of \ztwo-commutative
hermitian algebras, and each $\O^F(\cdot)$ becomes a hermitian
$\O$-module sheaf.

If the odd part of $E$ is finitedimensional, we get
for $U\seq E\seven$ a canonical isomorphism
\Beqn E1FinDim
 \O_E(U) \cong \An(U,\mathbb C)\otimes \opn S {E_{\odd,\mathbb C}}^*
\Eeq
where $\opn S {E\sodd}^*$ is the symmetric algebra over the complexified
dual of the odd part of $E$ in the supersense, i. e. the exterior algebra
in the ordinary sense.

Also, the Propositions 3.5.1, 3.5.2  from \cite{[CMP1]} carry over:

\begin{prp}
(i) Suppose that $u\in\O^F(U),\quad u_e=0$ for
some $e\in U$, and that $U$ is connected. Then $u = 0$.

(ii) Every Banach-valued analytic power series defines a "function
element": Assume that $F$ is Banach, and let be given
an element $v\in\P(E,p;F)$, $e\in E\seven$. Setting
\Beq
 U := e + \{e'\in E\seven:\quad p(e')<1\}
\Eeq
there exists a unique $F$-valued superfunctional $u\in\O^F(U)$ with
$u_e=v$. Explicitly, it is given by
\Beqn DefOfu
 u_{e+e'}:= \t_{e'}v
\Eeq
for $p(e')<1$.

(iii) If $v\in\P_{\opn pol }(E;F)$
is an analytic polynomial where $E,F$ may be arbitrary \ztwo-lcs
there exists a unique global $F$-valued superfunctional $u\in\O^F(E\seven)$
with $u_0=v$. Explicitly, it is given by
\eqref{DefOfu} again.
\qed\end{prp}

We conclude with a useful criterion for a map \eqref{SFctCand} to be
a superfunction.

Fixing a \ztwo-lcs $F$, we call a subset $V\seq F^*\seven \cup F^*\sodd$
{\em strictly separating} iff there exists a defining
system of seminorms $C\seq\CS(F)$ such that for all $p\in C$,
$f\in F$ with $p(f)\ne0$ there exists some $f^*\in V$ which
satisfies $\br|f^*|\le Kp$  with some $K>0$ and $f^*(f)\ne0$ (in
other words, we require that for every $p\in C$ the set of that
$f^*\in V$ which factorize through $\hat F_p$ separate that
Banach space).

Obviously, every strictly separating $V$ is separating (i. e.
it separates the points of $F$). Conversely, if $F$ is a \ztwo-Banach
space then every separating $V\seq F^*$ is strictly separating.
On the other hand, if $F$ carries the weak topology $\sigma(F,F^*)$
then $V\seq F^*\seven \cup F^*\sodd$ is strictly separating iff it
generates $F^*$ algebraically.

\begin{prp}\label{FQt:StrictSep}
Let $U\seq E\seven$ be open, let $F$ be a \ztwo-lcs, let
$V\seq F^*$ be strictly separating, and let be given a map
\eqref{SFctCand} such that for all $f^*\in V$, the map
$e\mapsto\br<f^*,u_e>\in \P(E;\mathbb C)$
is an element of $\O(U)$. Then \eqref{SFctCand} is an element of $\O^F(U)$.
\end{prp}

\begin{proof}
First we treat the case that $F$ is a Banach space. Fix $e\in E$ and choose
$p\in\CS(E)$ with $u_e\in\P(E,p;F)$. For $e'\in E\seven$ with $p(e')<1$,
we get $\br<f^*,u_{e+e'}-t_{e'}u_e>=0$ for all $f^*\in V$; since $V$  separates
$F$ we get $u_{e+e'}=t_{e'}u_e$, and the assertion follows.

Now let $F$ be an arbitrary \ztwo-lcs. It follows from the Banach case and the
definition of being strictly separating that for every $p\in C$ the assignment
$e\mapsto i_p\circ u_e$ is an element of $\O^{\hat F_p}(U)$.
The assertion follows.
\end{proof}

Surprisingly, the conclusion becomes false if $V$ is only supposed to
be separating. For a counterexample, cf. \cite[4.1]{[IS2]}.

\subsection{Superfunctions and ordinary functions}\label{SFctAnFu}
Before proceeding, we recall from \cite{[CMP1]} the definition of
real-analytic maps between locally convex spaces:

\begin{dfn} Let $E, F$ be real locally convex vector spaces (no
\ztwo grading), and let $U\seq E$ be open. A map $v: U\to F$ is
called {\em real-analytic} iff there exists a family $(u_k)_{k\geq 0}$
of continuous maps
\Beq
 u_k: U\times\underbrace{E\times\cdots\times E}_{\text{$k$ times}} \to F
\Eeq
which are symmetric and multilinear in the last $k$ arguments such that
for each $q\in\CS(F)$,\ \ $e\in U$ there exists some open $V\owns 0$
with $e+V\seq U$ such that for all $e'\in V$
\Beq
 \sum_{k\geq 0} i_q(u_k(e,e',\dots,e')) = i_qv(e+e')
\Eeq
with absolute convergence in the Banach space
$\hat F_q$, which is uniform in $e'$, i.~e.
\Beq
 \sup_{e'\in V}\sum_{k\geq k'} q(u_k(e,e',\dots,e')) \to 0
\Eeq
for $k'\to\infty$.
\end{dfn}

Of course, the $u_k$ are the Gateaux derivatives of $v$.

We denote by $\An(U,F)$ the set of all real-analytic maps
from $U$ to $F$.

\Brm
(1) Of course, if $F$ is Banach then it is
sufficient to check the condition only for $q$ being the
original norm, i.~e., $i_q$ is the identity map. On
the other hand, there exist (somewhat pathological) examples
where it is impossible to find a common $V$ for all $q\in\CS(F)$.
Cf. \cite{[IS1]} for a discussion and for the connection
of real-analytic maps with holomorphic maps between complex
locally convex spaces.

(2) The assignment $U\mapsto\An(U,F)$ is a sheaf of vector
spaces on $E$. Also, the class of real-analytic maps is closed
under pointwise addition and tensor product as well as under composition.
\Erm

As in \CMPref{3.6}, every $u\in\O^F(U)$ determines a real-analytic
{\em underlying function}
\Beq
 \tilde u: U\to F_{\mathbb C},\qquad e\mapsto \tilde u(e):=u_e[0].
\Eeq
In terms of the Taylor expansion $u_0\in\P(E,p;F)$ at the origin,
one has the explicit formula
\Beq
 \tilde u(e) = \sum_{k\ge0}
 (u_0)_{(k|0)}(\underbrace{e,\dots,e}_{\text{$k$ factors}})
\Eeq
valid for $e\in E\seven$,\ \ $p(e)<1$; if $u$ is polynomial then this formula
holds for all $e\in E\seven$.

\CMPref{Thm. 3.6.2} carries over:

\begin{thm}\label{CmpOAn}
(i) The map
\Beqn O2An
 \O^F(U)\to\An(U,F_{\mathbb C}),\qquad u\mapsto\tilde u,
\Eeq
is surjective, i. e. every real-analytic map is the underlying
functional of some superfunctional. Moreover, \eqref{O2An} turns
products into pointwise tensor products, and it turns hermitian conjugation
into conjugation within $F_{\mathbb C}$:
\Beq
 \widetilde{(u\otimes v)}(e) = \tilde u(e)\otimes\tilde v(e),\qquad
 \tilde{(\cj u)} (e) = \cj{\tilde u(e)}.
\Eeq
Also, it turns even directional derivative into the
Gateaux derivative: For $e'\in E\seven$ we have
\Beq
  \widetilde{\partial_{e'} u}(e) = D\tilde u(e,e').
\Eeq

(ii) If $E\sodd = 0$ then \eqref{O2An} is bijective. Explicitly, for a given
real-analytic map $v:U\to F_{\mathbb C}$ the corresponding superfunctional
$u\in\O^F(U)$ is given by $(u_e)_{(k|l)} = 0$ for $l\ne0$,
\Beq
 (u_e)_{(k|0)}(e_1,\dots,e_k) = D^kv(e,e_1,\dots,e_k)
\Eeq
where $D^kv$ is the $k$-th Gateaux derivative of $v$.
\qed\end{thm}

We will denote by $\M^F(\cdot):= \O^F(\cdot)\sevR$
the even, real part of $\O^F$.

The material of \CMPref{3.7} carries over:

\begin{prp}
Given $v\in\M_E^{E'}(U)$,\ \ $u\in\O_{E'}^F(U')$ with
$\opn Im \tilde v\seq U'$, the map
\Beq
 U\to \P(E;F),\quad e\mapsto (u\circ v)_e:=u_{\tilde v(e)}[v_e - \tilde v(e)]
\Eeq
is an element $u\circ v = u[v]\in\O^F(U)$ called the
{\em composition of $u$ with $v$}.
\qed\end{prp}

The standard coordinate, which was introduces in \ref{Insts}
only as power series, now globalizes by assertion (iii) of the
Proposition to a superfunction
$x = x_E\in\M^E(E\seven)$ which we call again the {\em standard coordinate}.

As in \CMPref{3.5}, it is often advisable to distinguish the "expansion
parameter" $x^e$ at a point notationally from the standard coordinate $x$.

\subsection{Superdomains} Given a complete \ztwo-lcs, we can assign to it
the hermitian ringed space (cf. \ref{HermStuff})
\Beq
 \L(E):=(E\seven,\O);
\Eeq
we call it the {\em linear supermanifold modelled over $E$}. The assignment
$E\mapsto\L(E)$ is the superanalogon of the usual procedure of viewing
a vector space as manifold.

Also, we call every open subspace $U=(\Space(U),\O|_U)$ of it a
{\em superdomain modelled over E}, and we write (symbolically)
$U\seq\L(E)$.

A {\em morphism of superdomains}
\Beqn MorOfSDs
 \phi:(U,\O_U)\to(V,\O_V)
\Eeq
which are modelled over \ztwo-lcs $E,\ F$, respectively,
is a morphism of hermitian ringed spaces which satisfies the following
condition:

There exists an element $\hat\phi\in\M^F(U)$ such that

(1) the underlying map of $\phi$ is given by the underlying function
$\tilde{\hat\phi}:U\to F$ of $\hat\phi$, and

(2) the superfunction pullback is given by composition with
$\hat\phi$: For open $V'\seq V$,
\ \ $\phi^*:\O_V(V')\to\O_U(\phi^{-1}(V'))$ maps
\Beqn ActOfSFP
 u\mapsto u\circ\hat\phi.
\Eeq
Using the standard coordinates $x_U\in\M^E(U)$, $x_V\in\M^F(V)$, this
rewrites to
\Beqn ActOfSFPInCos
 u[x_V]\mapsto u[\hat\phi[x_U]].
\Eeq
We note that $\hat\phi$ is uniquely determined by $\phi$ since for all
$f^*\in F^*$ we have $\left<f^*,\hat\phi\right> = \phi^*(f^*)\in\O(U)$
(on the r.h.s., we view $f^*$ as linear superfunction on $V$).

\Brmn MorOfFinSD
We note that in case that both $E,F$ are finitedimensional,
the additional requirement
of the existence of $\hat\phi$ is automatically satisfied and hence
redundant; that is, every morphism \eqref{MorOfSDs} of hermitian ringed spaces
is a morphism of superdomains. Indeed, if $f_i$ is a basis of $F$ and
$f^i$ the left dual bases of $F^*$ then it follows from the hermitian,
real-analytic version of Cor. \ref{FinDimPbChar} that
$\hat\phi=\sum\phi^*(f^i)f_i$ satisfies our requirements. However, in the
general case, its lifting would allow "nonsense morphisms", and there
would be no analogon to Cor. \ref{FinDimPbChar}.
\Erm

Given a fixed morphism \eqref{MorOfSDs}, the formulas \eqref{ActOfSFP},
\eqref{ActOfSFPInCos} are applicable also to $G$-valued superfunctions
where $G$ is an arbitrary complete \ztwo-lcs. We get a pullback map
\Beqn GValPBMap
 \phi^*:\O^G_V\to\phi_*\O^G_U.
\Eeq
In particular, $\hat\phi$ is now simply given as the pullback of the
standard coordinate:
\Beq
 \hat\phi=\phi^*(x_V).
\Eeq
Conversely, given an element $\hat\phi\in\M^F(U)$,
it determines a unique superdomain morphism $\phi:(U,\O_U)\to\L(F)$.
Thus, we get a bijection of sets
\Beqn MFRepresents
 \M^F(U)\to\{ \text{morphisms\ \ }(U,\O_U)\to\L(F)\},\qquad
 \phi\mapsto\hat\phi.
\Eeq

Now, given \eqref{MorOfSDs} and another morphism of superdomains
$\psi:(V,\O_V)\to(W,\O_W)$ then the composite $\psi\circ\phi$ is a
morphism, too; we have
\Beq
 \widehat{\psi\circ\phi} = \phi^*(\hat\psi).
\Eeq
Hence the superdomains form a category.

\Brm
\relax\eqref{MFRepresents} shows that for each $F$, the cofunctor
$(U,\O_U)\to\M^F(U)$ is represented by the linear superspace
$\L(F)$; the universal element is just the standard coordinate
$x\in\M^F(\L(F))$.
\Erm

\label{LinearMors}
Fix  \ztwo-lcs' $E,F$. Every even, continuous map
$\alpha: E\to F$ yields a morphism of superdomains
\Beq
 \L(\alpha): \L(E)\to\L(F),\quad
 \widehat{\L(\alpha)}:=\alpha\circ x_E\in \M^F(\L(E)).
\Eeq
Of course, the underlying map is the restriction of $\alpha$ to $E\seven$
while the superfunction pullback is given by
\Beq
  \left(\L(\alpha)^*(u)_e\right)_{(k|l)}:
  \prod^kE\seven\times\prod^lE\sodd \too {\prod^{k+l}\alpha}
  \prod^kF\seven\times\prod^lF\sodd \too {(u_e)_{(k|l)}} \mathbb C
\Eeq
for $u\in\O(\alpha(U))$,\ \ $e\in U$,\ \ $k,l\ge0$.

We call any morphism $\phi$ of superdomains
$\L(E)\supseteq U\too\phi V\to\L(F)$ which is the restriction of some
$\L(\alpha)$ a {\em linear morphism}.

\subsection{Supermanifolds}\label{Smfs}
In principle, a supermanifold
(abbreviated smf) $X$ is a hermitian ringed space which
locally looks like a superdomain. However, such a definition would be
insufficient because we cannot guarantee a priori that the arising
transition map $(U,\O_U)\to(U',\O_{U'})$ between two local models is a
morphism of superdomains. Therefore we add to the structure an atlas
of local models. That is, we define a {\em supermanifold
$X =(\Space(X),\O_X,(c_i)_{i\in I})$
(modelled over the complete \ztwo-lcs $E$)} as
consisting of the following data:

 (1) a Hausdorff space $\Space(X)$,

 (2) a sheaf of hermitian algebras $\O=\O_X$ on it,

 (3) a family of {\em charts}, i. e. of isomorphisms of hermitian ringed
spaces
\Beqn TheCharts
 c_i:(U_i,\O_X|_{U_i})\to(c_i(U_i),\O)
\Eeq
where $(U_i)_{i\in I}$ is an open covering of $X$, and the
$(c_i(U_i),\O)$ are superdomains modelled over $E$; in particular,
$c_i(U_i)\seq E\seven$ is open.

Setting $U_{ij}:=U_i\cap U_j$, there arise transition morphisms
between the local models,
\Beqn TheTransMors
 g_{ij}:=c_ic_j^{-1}:(c_j(U_{ij}),\O)\to(c_i(U_{ij}),\O)
\Eeq
and we require them to be morphisms of superdomains.

\Brm
For a formulation without charts cf. \cite{[IS1]}. For
the purposes of the "synthetic" approach presented here, which
explicitly indicates the form of morphisms to be allowed, the
point distributions of \cite{[IS1]} are not needed at all.

One could also use an approach via Douady's "functored spaces"
(cf. \cite{[Dou]}) in order to save charts.
\Erm

\medskip

The simplest example arises when there is just one chart, i. e.
$\Space(X)$ is effectively an open set $X\seq E\seven$. We then
call $X$ a {\em superdomain}; for $X=E\seven$ we get the linear
supermanifold $\L(E)$.

Fix an smf $X$. In order to define the {\em sheaf $\O^F$ of
$F$-valued superfunctions on $X$} we follow the usual "coordinate
philosophy" of differential geometry: Working on an open $U_i\seq X$
means actually to work on the superdomain $c_i(U_i)\seq\L(E)$;
the cocycle $(g_{ij})$ tells us how to pass from $U_i$ to $U_j$.

Thus, for open $U\seq X$, an element of $\O^F_X(U)$ is a family
$(u_i)_{i\in I}$ of elements
\Beqn ElOFAsFam
 u_i\in\O^F(c_i(U\cap U_i))
\Eeq
which satisfies the compatibility condition
$g_{ij}^*(u_i)=u_j$ on $c_j(U\cap U_{ij})$.

To make that more explicit, let $x_i\in\M^E(c_i(U_i))$ be the
standard coordinate. Then $u$ is given on $U\cap U_i$ by a
superfunction \eqref{ElOFAsFam} in the sense of \ref{SFcts}, $u_i=u_i[x_i]$,
and compatibility now means
\Beq
 u_j[x_j]=u_i[\hat g_{ij}[x_j]]
\Eeq
on $c_j(U\cap U_{ij})$. Here
$\hat g_{ij}=g_{ij}^*(x_i)\in\M(c_j(U_{ij}))$.

$\O^F_X$ is a sheaf of hermitian vector spaces.

For fixed $j$ we get an isomorphism of sheaves
\Beq
 \O^F_X|_{U_i}\to c_j^{-1}(\O^F_E),\qquad
 u\mapsto c_j^{-1}(u_i).
\Eeq
Also, we get an isomorphism
\Beq
\O=\O^{\mathbb R},\qquad
v\mapsto\left((c_i^{-1})^*(v)\right),
\Eeq
and we can identify both sides.

The product of \ref{FPSer} globalizes in an obvious way to a bilinear
map
\Beqn TheDirProd
 \O^F(U)\times\O^G(U)\to\O^{F\otimes G}(U).
\Eeq
Now, given a bilinear map $\alpha:F\times G\to H$ with a third
\ztwo-lcs $H$, we can compose \eqref{TheDirProd} with the map induced by
$\alpha:F\otimes G\to H$ to get a bilinear map
\Beqn TheBilMap
 \O^F(U)\times\O^G(U)\to\O^H(U).
\Eeq
It is natural to denote the image of $(u,v)$ simply by $\alpha(u,v)$.

\subsection{Morphisms of smf's}\label{FQt:MorOfSmf}
A {\em morphism of smfs} $\mu:(X,\O_X)\to(Y,\O_Y)$ is a morphism of
hermitian ringed spaces which is "compatible with the chart structure".
That is, we require that if
\Beq
 c_i:(U_i,\O_X|_{U_i})\to(c_i(U_i),\O),\qquad
 d_j:(V_j,\O_Y|_{V_j})\to(d_j(V_j),\O)
\Eeq
are charts on $X$, $Y$ then the arising composite morphism of hermitian
ringed spaces
\Beqn MorOfRSpac
 (c_i(\mu^{-1}(V_j)\cap U_i),\O) \too{c_i^{-1}}
 X \too\mu Y \too{d_j} (d_j(V_j),\O)
\Eeq
is a morphism of superdomains.

It is easy to check that the supermanifolds form a category;
the superdomains now form a full subcategory.

Fix a morphism $\mu:X\to Y$. Globalizing \eqref{GValPBMap}, we define
the {\em pullback of $\O^G$},
\Beq
 \mu^*:\O^G_Y\to\mu_*(\O^G_X)
\Eeq
as follows: Let $v\in\O^G_Y(V)$, i. e. $v=(v_j)$ with
$v_j=v_j(x_j)\in\O^G(d_j(V\cap V_j))$. We put together the pullbacks
of the $v_i$ under the superdomain morphisms \eqref{MorOfRSpac}: Fixing
$i$, the elements
\Beq
 (d_j\mu c_i^{-1})^*(v_i)\in\O^G(c_i(\mu^{-1}(V\cap V_j)\cap U_i))
\Eeq
with running $j$ fit together to an element
\Beq
 u_i\in\O^G(c_i(\mu^{-1}(V)\cap U_i));
\Eeq
and $\mu^*(v):=(u_i)\in\O^G(\mu^{-1}(V))$ is the pullback wanted.

The bijection \eqref{MFRepresents} globalizes to smfs, yielding the
infinite-dimensional version of Cor. \ref{FinDimPbChar}:

\begin{thm}\label{CoordChar}
A morphism of an smf $X$ into a linear smf $\L(E)$ is characterized
by the pullback of the standard coordinate $x_E$. That is, we have
a bijection
\Beq
 \opn Mor (X,\L(E))\to\M^E(X),\qquad \mu\mapsto\hat\mu:=\mu^*(x_E).
\Eeq
\qed\end{thm}

This is the superanalogon of the (tautological) non-super fact that a
real-analytic function on a manifold $M$ with values in a vector space
$F$ is the same as a real-analytic map $M\to F$ (cf. also
\cite[Thm. 3.4.1]{[IS1]}).

\subsection{Supermanifolds and manifolds}\label{SmfsAMfs}
We now turn to the relations of smfs with ordinary real-analytic manifolds
modelled over locally convex spaces; since we defined in \ref{SFctAnFu}
real-analytic maps, this notion makes obvious sense.

Fix an smf $X$ and charts \eqref{TheCharts}.
It follows from \ref{SFctAnFu} that the underlying map
$\tilde g_{ij}:c_j(U_{ij})\to c_i(U_{ij})$
of each transition morphism \eqref{TheTransMors} is real-analytic; therefore
the maps $\tilde c_i:U_i\to E\seven$ equip $\Space(X)$ with the structure of
a real-analytic manifold with model space $E\seven$ which we call the
{\em underlying manifold} of $X$ and denote by $\tilde X$.  Assigning
also to every smf morphism its underlying map we get a functor
\Beqn Underlfunctor
 \Smfs\to \Mfs,\qquad X\mapsto \tilde X
\Eeq
from the category of supermanifolds to the category of manifolds.

We want to construct a right inverse to \eqref{Underlfunctor}, i. e. we want
to assign to every manifold $Y$ a supermanifold, denoted $Y$ again, the
underlying manifold of which is $Y$ again.
First we note that given an lcs $F$, we may view it as
\ztwo-lcs by $F\sodd:=0$. Let $\An_F(\cdot)$ be the
sheaf of all real-analytic $\mathbb C$-valued functions on $Y$ with
the obvious hermitian algebra structure.
We recall from  Thm. \ref{CmpOAn}.(ii) that we have an isomorphism
of hermitian algebra sheaves $\An_F(\cdot) \cong \O(\cdot)$.

Now let be given a real-analytic manifold $Y$ with model space
$F$. We want to assign to $Y$ a supermanifold
$Y=(\Space(Y),\O_Y,(c_i)_{i\in I})$ with model space $F$ (which
we view by $F=F\seven$ as \ztwo-lcs). Of course, we take the space $Y$
as underlying space, and the
sheaf $\O_Y(\cdot):= \An(\cdot,\mathbb C)$ of real-analytic
$\mathbb C$-valued functions on $Y$ with the obvious hermitian algebra
structure as structure sheaf. In order to get the needed atlas $(c_i)_{i\in
I}$,
we choose an atlas on the manifold $Y$ in the usual sense, i. e.
a collection of open injective maps $c'_i:V_i\to F$ where
$Y=\bigcup_{i\in I} V_i$ such that both $c'_i$ and
$(c'_i)^{-1}:c'_i(V_i)\to V_i$ are real-analytic. Using the usual
function pullback we can view the $c_i$ as morphisms of hermitian ringed spaces
\Beq
 c_i:(V_i,\O_Y|_{V_i}) = (V_i,\An_{V_i}(\cdot,\mathbb C)) \to
 (c'_i(V_i),\An_{c'_i(V_i)}(\cdot,\mathbb C)) = (c'_i(V_i),\O_F|_{c'_i(V_i)});
\Eeq
we claim that these morphisms equip $Y$ with an smf structure.
It is sufficient to show that given a real-analytic map $\phi:U\to U'$
with $U,U'\seq F$ open, the arising morphism of hermitian ringed spaces
\Beq
 \phi:(U,\O_F|_{U})\to(U',\O_F|_{U'})
\Eeq
is a morphism of superdomains. Indeed, recalling that $\M^F$
is just the sheaf of $F$-valued  analytic maps, we can
view $\phi$ as an element of $\M^F(U)$, and this is the element
required in the definition of superdomain morphisms.

One easily shows that this construction is functorial and provides a
right inverse
\Beqn RightInverse
 \Mfs\to \Smfs
\Eeq
to \eqref{Underlfunctor}. Henceforth, we will not make any notational
distinction between a manifold $Y$ and the corresponding smf. In particular,
any purely even locally convex space $E$ can be viewed as manifold, and hence
as smf $E$.

In fact, the functor \eqref{RightInverse} identifies the category of
real-analytic manifolds with the full subcategory of all those
supermanifolds the model space of which is purely even.

Now let $X$ again be an smf. For each \ztwo-lcs $F$, we get
by assigning to each $F$-valued superfunction its underlying function
a sheaf morphism
\Beqn GlobUnderlFct
 \O^F\to\An(\cdot,F_{\mathbb C}),\qquad u\mapsto \tilde u;
\Eeq
by Thm. \ref{CmpOAn}.(ii), it is surjective on every open piece of $X$
which is isomorphic with a superdomain. Also, \eqref{GlobUnderlFct}
restricts to a sheaf morphism $\M^F\to\An(\cdot,F)$.

Now, applying onto $X$ first the functor \eqref{Underlfunctor} and then the
functor \eqref{RightInverse}, we get another smf $\tilde X$, and,
specializing \eqref{GlobUnderlFct} to $F:=\mathbb R$, we get a sheaf morphism
$\O_X\to \An(\cdot,\mathbb C) = \O_{\tilde X}$, and hence a morphism of
hermitian ringed spaces
\Beqn MorRgdSp
 (\tilde X,\O_{\tilde X})\to(X,\O_X)
\Eeq
the underlying point map of which is the identity. In fact, one shows by
looking at the local model that \eqref{MorRgdSp} is an smf morphism
\Beqn CanEmb
  \tilde X \to X
\Eeq
which we call the {\em canonical embedding}. It is functorial again: For
a given morphism $\mu:X\to Y$ of smfs we get a commutative diagram
\Bcd
  X        @>\mu>>        Y\nt
  @AAA                    @AAA \nt
  \tilde X @>\tilde \mu>> \tilde Y.
\Ecd

\Brm
(1) Of course, the term "canonical embedding" is slightly abusive as long as
the notion of embedding for smfs has not been defined.

(2) The functoriality of \eqref{CanEmb} is somewhat  deceptive
because it does not extend to {\em families} of morphisms:
Given a $Z$-family of morphisms from $X$ to $Y$, i. e. a morphism
$\mu: Z\times X\to Y$ (cf. below for the definition of the product),
there is no natural way to assign to it
$Z$-family of morphisms from $\tilde X$ to $\tilde Y$, i. e.
a morphism $Z\times \tilde X\to \tilde Y$. All what one has is
$\tilde\mu: \tilde Z\times \tilde X\to \tilde Y$, which is only a
$\tilde Z$-family.
\Erm

\subsection{Cocycle description; products}\label{CocycDescr}
Let us describe supermanifolds in terms of cocycles of superfunctions:
Assume that we are given the model space $E$ and a manifold
$\tilde X$ with model space $E\seven$ together with charts
\Beqn UndCh
 \tilde c_i:U_i\to \tilde c_i(U_i)\seq E\seven.
\Eeq
Let $\tilde g_{ij}:=\tilde c_i\tilde c_j^{-1}$; thus
$\tilde g_{ij}\circ\tilde g_{jk}= \tilde g_{ik}$
on $\tilde c_k(U_i\cap U_j\cap U_k)$.

Now we want to make $\tilde X$ the underlying manifold of a supermanifold
$X$ which is on each $U_i$ isomorphic to a superdomain in $\L(E)$
such that the maps \eqref{UndCh} are the underlying maps of corresponding
charts on $X$. For this purpose, we have to choose a family
$(\hat g_{ij})_{i,j\in I}$ of elements
$\hat g_{ij}\in\M^E(\tilde c_j(U_{ij}))$
with underlying function $\tilde g_{ij}$ and such that
$\hat g_{ii} = x_E$,
\Beq
 \hat g_{ij}\circ \hat g_{jk}=\hat g_{ik}
\Eeq
on $\tilde c_k(U_i\cap U_j\cap U_k)$.

For constructing an smf out of these data, we adapt the gluing prescription
given in \ref{Smfs} to produce a hermitian algebra sheaf $\O_X$:
For open $U\seq X$, an element of $\O_X(U)$ is a family
$(u_i)_{i\in I}$ of elements $u_i\in\O_{\L(E)}(\tilde c_i(U\cap U_i))$
which satisfies $u_i\circ\hat g_{ij}=u_j$ on $\tilde c_j(U\cap U_{ij})$.
Moreover, we construct a chart
\Beq
 c_i:(U_i,\O_X|_{U_i})\to(\tilde c_i(U_i),\O_{\L(E)}|_{\tilde c_i(U_i)})
\Eeq
by taking, of course, $\tilde c_i$ as underlying map and
$u \mapsto (u\circ\hat g_{ij})_{j\in I}$ as the superfunction pullback.
In particular, we get the local coordinate $x_i:=c_i^*(x_E)$, so that
\Beqn CoTrans
 x_j = g_{ji}\circ x_i.
\Eeq
It is clear that $X :=(\tilde X,\O_X,(c_i)_{i\in I})$ is the smf wanted;
we call $(\hat g_{ij})$ the {\em defining cocycle} of $X$.

\Brm
Given the data $E$, $\tilde X$, $\tilde c_i:U_i\to \tilde c_i(U_i)\seq E\seven$
as above, one obvious lifting $\tilde g_{ij}\mapsto \hat g_{ij}$ of the
transition maps does always exist;
it is easily seen to produce the product smf $\tilde X\times\L(E\sodd)$
(cf. below for the definition).
\Erm

We now show that the category of smfs has finite products:

\begin{prp}
Given smfs $X$, $Y$ there exists
an smf $X\times Y$ together with morphisms $\pr_X:X\times Y\to X$,
$\pr_Y:X\times Y\to Y$ such that the following universal property is
satisfied:

Given an smf $Z$ and morphisms $\alpha: Z\to X$,
$\beta: Z\to Y$, there exists a unique morphism $\zeta: Z\to X\times Y$
such that $\pr_X\circ\zeta=\alpha$,\ \ $\pr_Y\circ\zeta=\beta$.

We call $X\times Y$ the {\em product of $X$ and $Y$}, and
$\pr_X,\pr_Y$ the {\em projections}.
\end{prp}

\begin{proof} We first treat the case that $X,Y$ are superdomains
$U\seq\L(E)$,\ \  $V\seq\L(F)$. Of course, the product sought is just the
superdomain $U\times V\seq \L(E\oplus F)$ the underlying domain of which is
$\Space(U)\times\Space(V)$; we now define the projection
$\pr_U: U\times V \to U$ to be the linear morphism (cf. \ref{LinearMors})
induced by the projection $E\oplus F\to E$.\ \
$\pr_V$ is given quite analogously.

Now, given $\alpha: Z\to U$, $\beta: Z\to U$, we get elements
$\hat\alpha\in\M^E(Z)\seq \M^{E\oplus F}(Z)$,\ \
$\hat\beta\in\M^F(Z)\seq \M^{E\oplus F}(Z)$, and it is easy to see that
the morphism $\zeta:Z\to U\times V$ given
by $\hat\zeta:=\hat\alpha+\hat\beta$ implements the universal property wanted.

Turning to the case of arbitrary smfs, we use the cocycle construction of
above: Let be given smfs $X$, $Y$ with model spaces $E,\ F$,
and charts $c_i$ on $U_i$, $d_j$ on $V_j$, respectively. We get cocycles
\Beq
 \hat g_{ik}:=\widehat{c_ic_k^{-1}} \in\M^E(\tilde c_k(U_{ik})),\quad
 \hat h_{jl}:=\widehat{d_jd_l^{-1}} \in\M^F(\tilde d_l(V_{jl})).
\Eeq
Now $\tilde X\times\tilde Y$ is a manifold with charts
$\tilde c_i\times\tilde d_j:\Space(U_i)\times\Space(V_j)
\to E\seven\oplus F\seven$ and transition functions
\Beq
 \tilde c_i\times \tilde d_j(\tilde c_k\times\tilde d_l)^{-1}=
 \tilde g_{ik}\times\tilde h_{jl}.
\Eeq
Using the lifting mechanism of above, we can lift $\hat g_{ik},\;\hat h_{jl}$
to elements $\hat g_{ik},\;\hat h_{jl}\in\M^{E\oplus F}
 (\tilde c_k(U_{ik})\times\tilde d_l(U_{jl}))$, and  we take
$\hat g_{ik}\hat h_{jl}$ as the defining cocycle of the supermanifold
$X\times Y$ with underlying manifold $\tilde X\times\tilde Y$.
This smf has charts $e_{ij}:(\Space(U_i)\times\Space(V_j),\O_{X\times Y})
 \to c_i(U_i)\times d_j(V_j)$; the compositions
\Beq
 \Space(U_i)\times\Space(V_j)\too{\pr_1\circ e_{ij}} c_i(U_i) \too{c_i^{-1}}
U_i \too\seq X
\Eeq
agree on the overlaps and therefore glue together to an smf morphism
$\pr_X: X\times Y\to X$; one defines $\pr_Y$ analogously.

Using the superdomain case and some chart juggling, it follows that our
requirements are satisfied.
\end{proof}

\Brm
The category of smfs has also a terminal object $P=\L(0)$ which is
simply a point.
\Erm

\subsection{Comparison with finite-dimensional Berezin smfs}\label{CompFC}
Let $X$ be an smf with model space $E$, and $F$ be any \ztwo-lcs. We
get an embedding of sheaves
\Beqn EmbOFOF
 \O(\cdot)\otimes F \to \O^F(\cdot),\quad
 u\otimes f\mapsto (-1)^{\br|f|\br|u|} f\cdot u
\Eeq
where, of course, on every coordinate patch
\Beq
 ((f\cdot u)_e)_{(k|l)}(e_1,\dots,e_k,e'_1,\dots,e'_l) =
 f\cdot (u_e)_{(k|l)}(e_1,\dots,e_k,e'_1,\dots,e'_l)
\Eeq
for $e_1,\dots,e_k\in E\seven$,\ \  $e'_1,\dots,e'_l\in E\sodd$.

For arbitrary $F$, \eqref{EmbOFOF} is far away from being isomorphic; in fact,
the image consists of all those $u\in\O^F(\cdot)$ for which there exists
locally a finite-dimensional \ztwo-graded subspace $F'\seq F$ such that
$u\in\O^{F'}(\cdot)$.

It follows that if $F$ is itself finite-dimensional then
\eqref{EmbOFOF} is an isomorphism; this is the reason why the sheaves
$\O^F(\cdot)$ become important only in the infinite-dimensional context.

{}From Thm. \ref{CoordChar} we now get the usual characterization of
morphisms by coordinate pullbacks (cf. Thm. \ref{FinDimCoords}).

\begin{cor}
Let $X$ be an smf, let $F$ be a finite-dimensional \ztwo-graded vector
space, and let $f_1,\dots,f_k,f'_1,\dots,f'_l\in F^*\seq \O(\L(F))$ be a
basis of the dual $F^*$.

Given elements $u_1,\dots,u_k\in\M(X)$,\ \
$v_1,\dots,v_l\in\O(X)_{\odd,\mathbb R}$ there exists a unique smf morphism
$\mu: X\to \L(F)$ with the property $\mu^*(f_i)=u_i$, $\mu^*(f'_j)=v_j$
($i=1,\dots,k$, $j=1,\dots,l$). It is given by
\Beq
 \hat\mu = \sum_{i=1}^k f^i\cdot u_i + \sum_{j=1}^l (f')^i\cdot v_i
 \in \M^F(X)
\Eeq
where $f^1,\dots,f^k,(f')^1,\dots,(f')^l\in F$ is the left dual basis.
\qed\end{cor}

We now turn to the comparison of our smf category with
with the category of finite-dimensional real-analytic Berezin smfs.
Adapting the definition given in \ref{HermStuff} to our real-analytic
situation, we define a {\em (hermitian) real-analytic Berezin smf}
$X$ as a hermitian ringed space $(\Space(X),\O)$, where $\Space(X)$ is required
to be Hausdorff, and which is locally isomorphic to the model space
\Beqn HRABerStrShf
 \O(\cdot)=\An(\cdot,\mathbb C)\otimes_{\mathbb C}\mathbb C[\xi_1,\dots,\xi_n]
\Eeq
where $\xi_1,\dots,\xi_n$ is a sequence of Grassmann variables.
The hermitian structure on \eqref{HRABerStrShf} is given by
\eqref{HermConjLaw} again.

Note that we have silently dropped the usual requirement of paracompactness
since we do not know an infinite-dimensional generalization of it.

Now, given an smf $(\Space(X),\O_X,(c_i)_{i\in I})$ in the sense of \ref{Smfs}
whose model space is finite-di\-men\-sio\-nal, it follows from the
isomorphism \eqref{E1FinDim} that we only need to forget the charts
to get a real-analytic Berezin smf
$(\Space(X),\O_X)$. Conversely, given such a real-analytic Berezin smf, we
can take as atlas e. g. the family of all isomorphisms
$c: U\to \L(\mathbb R^{m|n})$ where $U\seq X$ is open; since, as already
observed in Rem. \ref{MorOfFinSD}, the morphisms
of finitedimensional superdomains "care for themselves", this is OK.

\begin{cor}
We have an equivalence of categories between
\begin{itemize}
\item
 the full subcategory of the category of smfs formed by the
 smfs with finite-dimensional model space, and
\item
 the category of hermitian real-analytic Berezin smfs,
\end{itemize}
which acts on objects as
\Beq
 (\Space(X),\O_X,(c_i)_{i\in I})\mapsto (\Space(X),\O_X).
\Eeq
\qed\end{cor}

Finally, we consider a (fairly simple) functional calculus for scalar
superfunctions: Fix an smf $X$. Every element $f\in\M(X)$ can be
interpreted as a morphism $f:X\to\L(\mathbb R)=\mathbb R$
(on the other hand, any odd superfunction $f$ gives
rise to a morphism $\Pi f:X\to\L(\Pi\mathbb R)$, but we will not use that).
Now if $F:U\to\mathbb R$ is an analytic function on an open set $U\seq\mathbb
R^n$,
and if $f_1,\dots,f_n\in\M(X)$ then one can make sense of
the expression $F(f_1,\dots,f_n)$ provided that
\Beq
 (\tilde f_1(x),\dots,\tilde f_n(x))\in U
\Eeq
for all $x\in X$.
Indeed, in that case, $F(f_1,\dots,f_n)\in\M(X)$ is the superfunction
corresponding to the composite morphism
\Beq
 X \too{(f_1,\dots,f_n)} U \too F \mathbb R.
\Eeq
One easily shows:

\begin{cor}
(i) If
$F = F(z_1,\dots,z_n) =
 \sum_{|\alpha|\le N}c_\alpha z_1^{\alpha_1}\cdots z_n^{\alpha_n}$
is a polynomial then
$F(f_1,\dots,\allowbreak f_n)\allowbreak =
 \sum_{|\alpha|\le N}c_\alpha f_1^{\alpha_1}\cdots f_n^{\alpha_n}$.

(ii) For $i=1,\dots,k$ let $U_i$ be open in $\mathbb R^{n_i}$, let $V$ be
open in $\mathbb R^k$. Let $F_i\in\An(U_i,\mathbb R)$, $G\in\An(V,\mathbb R)$,
suppose that $(F_1(z_1),\dots,F_k(z_k)\in V$ for all
$(z_1,\dots,z_k)\in U_1\times\dots\times U_k$, and set
\Beq
 H(z_1,\dots,z_k):=G(F_1(z_1),\dots,F_k(z_k))
\Eeq
for all $(z_1,\dots,z_k)\in U_1\times\cdots\times U_k$. Let
$f_i=(f_i^1,\dots,f_i^{n_i})\in\prod^{n_i} \M(X)$ for $i=1,\dots,k$,
and suppose that the image of $f_i:X\to\mathbb R^{n_i}$ lies in $U_i$.
Then we have the identity
\Beq
 H(f_1,\dots,f_k):=G(F_1(f_1),\dots,F_k(f_k))
\Eeq
in $\M(X)$.
\qed\end{cor}

\begin{cor}\label{InvOfSfct}
An even scalar superfunction
$f\in\O(X)\seven$ is invertible iff $\tilde f(x)\ne0$ for all $x\in X$.
\end{cor}

\begin{proof}
The condition is clearly necessary. Set $U:=\mathbb R\setminus\{0\}$,
\ \ $F:U\to U$,\ \ $c\mapsto c^{-1}$. We have $\cj ff\in\M(X)$ for
$f\in\O(X)\seven$, and if $f$ satisfies our condition then we may form
$f^{-1}:=\cj f\cdot F (\cj ff)$; using the Cor. above we get $f^{-1}f=1$.
\end{proof}

Thus, each stalk $\O_x$ of the structure sheaf is a (non-Noetherian)
local ring.

Also, $\exp(f)$ is defined for any $f\in\M(X)$, and we have the identity
$\exp(f+g)=\exp(f)\exp(g)$.

\subsection{Sub-supermanifolds}
Essentially, we will follow here the line of \cite{[Lei1]}.

Let be given smf's $X,Y$ with model spaces $E,F$, respectively, and
an smf morphism $\phi: X\to Y$.
We call $\phi$ {\em linearizable} at a point $x\in X$ iff, roughly spoken,
it looks at $x$ like a linear morphism (cf. \ref{LinearMors}), i.e. iff
there exist neighbourhoods $U\owns x$,\ \ $V\owns \tilde\phi(x)$, superdomains
$U'\seq\L(E)$, $V'\seq\L(F)$, and isomorphisms
$i_U: U\to U'$, $i_V: V\to V'$ such that the composite
$\phi':= i_V\phi(i_U)^{-1}: U'\to V'\seq\L(F)$
is a linear morphism $\L(\alpha)$. We then call
$\alpha: E\to F$ the {\em model map} of $\phi$ at $x$.
(We note that the model map is uniquely determined up to automorphisms of
$E,F$ since it can be identified with the tangent map
$\T_xX\to\T_{\tilde\phi(x)}Y$; cf. \cite{[IS2]}.)

We call an even linear map $F\to E$ of \ztwo-lcs a {\em
closed embedding} iff it is injective, its image is closed, and
the quotient topology on the image is equal to the topology induced
by the embedding into $F$. We call $F\to E$ a {\em split embedding}
iff it is a closed embedding, and there exists a closed subspace
$E'\seq E$  such that $E=E'\oplus F$ in the topological sense.
We recall that this is equivalent with the existence of a linear
continuous projection $E\to F$.

Thus, if $F$ is finite-dimensional and $E$ a
Fr\`echet \ztwo-lcs then every injective linear map $F\to E$
is a split embedding.

We call a morphism of smf's $\phi:X\to Y$ a {\em regular closed embedding} if
its underlying point map is injective with closed image, and
$\phi$ is linearizable at every point $x\in X$, with the model
map at $x$ being a closed embedding.

One easily shows that if $\phi:X\to Y$ is a regular closed embedding then
$\phi$ is a monomorphism in the sense of category theory, i.e. given
two distinct morphisms $Z\doublearrow X$ the composites
$Z\doublearrow X\too\phi Y$ are still different.

Two regular closed embeddings $\phi:X\to Y$,\ \ $\phi':X'\to Y$ are called
{\em equivalent} iff there exists an isomorphism $\iota: X\to X'$ with
$\phi'\circ\iota=\phi$; because of the monomorphic property, $\iota$ is
uniquely determined.

We call any equivalence class of regular closed embeddings into $Y$ a
{\em sub-supermanifold (sub-smf)} of $Y$, and we call the equivalence class
of a given regular closed embedding $\phi:X\to Y$ the {\em image} of $\phi$.
(Sometimes, by abuse of language, one calls $X$ itself a sub-smf, with
$\phi$ being understood.)

Finally, we call a sub-smf $\phi:Y\to X$ a {\em split sub-smf}
if its tangential map is everywhere a split embedding.

For example, the canonical embedding \eqref{CanEmb} makes $\tilde X$
a split sub-smf of $X$, with the model map being the embedding $E\seven\seq E$.

Note that the notion of a regular closed embedding is presumably in general not
transitive (although we do not know a counterexample). However, it is easy to
see that if $\phi:X\to Y$ is linearizable at $x$, and $Y\seq Z$ is a split
sub-smf then the composite $X \too\phi Y$ is linearizable at $x$, too. In
particular, the notion of a split sub-smf is transitive.

\label{CuttingOut}
Let be given an smf $X$ and a family of superfunctions $u_i\in\O^{F_i}(X)$,
\ \ $i\in I$, where $F_i$ are \ztwo-lcs. We call a sub-smf represented
by $\phi:Y\to X$ the {\em sub-smf cut out by the superfunctions $u_i$}
iff the following holds:

We have $\phi^*(u_i)=0$; moreover, if $\phi':Y'\to X$ is any other smf
morphism with $(\phi')^*(u_i)=0$ for all $i$ then there exists a morphism
$\iota:Z\to X$ with $\phi\circ\iota=\phi'$.

(Note that by the monomorphic property of $\phi$, $\iota$ is uniquely
determined.)

Of course, for a given family $(u_i)$, such a sub-smf needs not to exist;
but if it does, it is uniquely determined. Also, using the universal property
with $Y':=P$ being a point we get that if $Y$ exists then
$\tilde\phi: \Space(Y)\to\Space(X)$ maps $\Space(Y)$ homeomorphically onto
\Beq
 \{x\in\Space(X):\quad \widetilde{u_i}(x)=0\ \forall i\in I \}.
\Eeq
However, note that the superfunction pullback $\O_X\to\O_Y$ need not be locally
surjective; it is so only in a finite-dimensional context.

\Brm
One can show that appropriate formulations of the inverse and implicite
function theorems hold provided that the model spaces of the smf's
involved are Banach spaces.
\Erm

\subsection{Example: the unit sphere of a super Hilbert
space}\label{FQt:SUnSph}
As a somewhat academic example, let us consider the super variant of the
unit sphere of a Hilbert space: As in \cite{[HERM]}, a
{\em super Hilbert space} is a direct sum of two ordinary Hilbert spaces,
$H=H\seven\oplus H\sodd$; the scalar product will be
denoted by
\Beq
 H\times H\to\mathbb C,\qquad  (g,h)\mapsto\br<\cj g|h>
\Eeq
(the unusual notation helps to keep track of the action of the second
sign rule).

Let  $H_r$  denote  then  underlying real vector space of $H$.

Let $x\in\M^{H_r}(\L(H_r))$ denote the standard
coordinate. Specializing \eqref{TheBilMap} to the $\mathbb R$-bilinear pairing
\Beq
 H_r\times H_r\to\mathbb C,\qquad  (g,h)\mapsto\br<\cj g|h>
\Eeq
we get a bilinear pairing
\Beqn The<>Pair
 \O^{H_r}(\cdot)\times \O^{H_r}(\cdot)\to\O(\cdot).
\Eeq
Applying this to $(x,x)$ yields a superfunction denoted by
$\br\|x\|^2\in\M(\L(H_r))$ (the notation looks abusive, but can be
justified, cf. \cite[4.4]{[IS1]}). The Taylor series of $\br\|x\|^2$
at zero is given by $(\br\|x\|^2)_0 = u_{(2|0)} + u_{(0|2)}$ where
\Bal
 u_{(2|0)}: (H_r)\seven\times(H_r)\seven\to\mathbb C,\qquad
 &(g,h)\mapsto \br<\cj g|h>+\br<\cj h|g> = 2\cdot\Re\br<\cj g|h>,
 \\
 u_{(0|2)}: (H_r)\sodd\times(H_r)\sodd\to\mathbb C,\qquad
 &(g,h)\mapsto \br<\cj g|h>-\br<\cj h|g> = 2\cdot\Im\br<\cj g|h>.
\Eal
Of course, the underlying map is $(H_r)\seven\to\mathbb R$,\ \
$g\mapsto\br\|g\|^2$.

\begin{prp}\label{FQt:UnitSph}
Let $H$ be a super Hilbert space, and suppose that $H\seven\not=0$.
There exists a sub-smf $\mathbb S \too\epsilon \L(H_r)$ cut out
by the element $\br\|x\|^2-1\in\M(\L(H_r))$. Moreover, we have an
isomorphism
\Beqn H=RxS
 \iota: \mathbb S\times\mathbb R_+ \too\cong \L(H_r)\setminus0
\Eeq
where $\mathbb R_+=\{c\in\mathbb R:\ \ c>0\}$ is viewed by \ref{SmfsAMfs} as
smf.
We call $\mathbb S$ the {\em super unit sphere} of $\L(H_r)$.
\end{prp}

\begin{proof}
Looking at the situation with $Z=P$ being a point, we see that if
$\mathbb S$ exists its underlying manifold can be identified with the usual
unit
sphere $\tilde{\mathbb S}$ of $(H_r)\seven$.
In order to use stereographic projection, we fix an element
$h\in H\seven$, $\br\|h\|=1$; set
\Beq
 E:= \{\xi\in H_r: \quad\Re\br<\cj{\xi}|h> = 0 \}.
\Eeq
Note $(H\sodd)_r\subset E$; the real \ztwo-graded Hilbert space
$E$ will be the model space for $\mathbb S$. We get
an atlas of the manifold $\tilde{\mathbb S}$ consisting of two
homeomorphisms
\Beq
 \tilde c_\pm: \tilde{\mathbb S}\setminus\{\pm h\} \to E\seven,
\Eeq
\Beqn ActC
 \xi\mapsto \pm h +
 \frac 1{\pm\Re\br<\cj{\xi}|h> -1}(\pm h - \xi);
\Eeq
the inverse maps are ${\tilde c_{\pm}}^{-1}(\eta) =
 \eta\mapsto\pm h + \frac 2{1+\br\|\eta\|^2}(\eta \mp h)$.
The transition between the charts is
\Beq
 \tilde g:=\tilde c_+{\tilde c_-}^{-1}:
  E\seven\setminus\{0\}\to E\seven\setminus\{0\},\quad
 \eta\mapsto \eta/\br\|\eta\|^2.
\Eeq
In accordance with \ref{CocycDescr}, we lift $\tilde g$ to the superfunction
\Beq
 g[y]:=y/\br\|y\|^2\in\M^E(\L(E)\setminus\{0\}),
\Eeq
where $y\in\M^E(\L(E))$ is the standard coordinate.
We get an smf $\mathbb S$ with underlying manifold $\tilde{\mathbb S}$ and
model space $E$. The $\tilde c_\pm$ become the underlying maps of two charts
\Beq
 c_\pm: \mathbb S\setminus\{\pm h\} \to \L(E),
\Eeq
and we get coordinates
$y_\pm:=\widehat{c_\pm}={c_\pm}^*(y) \in\M^E(\mathbb S\setminus\{\pm h\})$;
from \eqref{CoTrans} we get
\Beqn TheSTrans
 y_- = y_+/\br\|y_+\|^2.
\Eeq
Set
\Beq
 e_\pm:= \pm h + \frac 2{1+\br\|y_\pm\|^2}(y_\pm \mp h)
 \in\M^{H_r}(\mathbb S\setminus\{\pm h\}).
\Eeq
Using \eqref{TheSTrans}, one computes that the restrictions of
$e_-,e_+$ onto $\mathbb S\setminus\{h,-h\}$ coincide;
hence we can define the smf morphism $\epsilon:\mathbb S\to\L(H_r)$ by
$\hat\epsilon|_{\mathbb S\setminus\{\pm h\}} = e_\pm$.
Of course, the underlying map  $\tilde\epsilon$ is the inclusion
$\tilde{\mathbb S}\subset H\seven$.
Using again the pairing \eqref{The<>Pair} one computes
\Beqn PullBIs0
 \epsilon^*(\br\|x\|^2-1) =
 \br<\cj{\hat\epsilon}|\hat\epsilon>^2-1 = 0.
\Eeq
Now we can define the morphism \eqref{H=RxS} by $\hat\iota := t\hat\epsilon$
(if being pedantic, one should write $\pr_1^*(t)\pr_2^*(\hat\epsilon)$
instead) where $t\in\M(\mathbb R_+)$ is the standard coordinate.

In order to show that \eqref{H=RxS} is an isomorphism, we construct
morphisms
\Beq
 \kappa_\pm: \L(H_r)\setminus(\pm \mathbb R_+ h)\to
 \mathbb R_+\times(\mathbb S\setminus\{\pm h\}),
 \quad
 {\kappa_\pm}^*((t,y_\pm)) = \Bigl(\br\|x\|,\ \
 \pm h + \frac 1{\pm\frac {\Re\br<\cj x|h>}{\br\|x\|} -1}
 (\pm h - \frac x{\br\|x\|})\Bigr)
\Eeq
where the superfunction
$\br\|x\| := \sqrt{\|x\|^2} \in\M(\L(H_r)\setminus0)$ is defined by
the functional calculus of \ref{CompFC}. Now $\kappa_+,\kappa_-$ coincide
on the overlap $\L(H_r)\setminus\mathbb R h$: One computes
\Beqn |Kap*y|
 \br\|{\kappa_+}^*(y_+)\|^2 =
 \frac {-\Re\br<\cj x|h>-\br\|x\|}{\Re\br<\cj x|h>-\br\|x\|}
\Eeq
and hence
\Beq
 {\kappa_+}^*(y_-) = {\kappa_+}^*\br(\frac 1{\br\|y_+\|^2}y_+) =
 \frac 1{\br\|{\kappa_+}^*(y_+)\|^2} {\kappa_+}^*(y_+)  =
 \frac {\Re\br<\cj x|h>-\br\|x\|}{-\Re\br<\cj x|h>-\br\|x\|} {\kappa_+}^*(y_+)
=
 {\kappa_-}^*(y_-).
\Eeq
Hence $\kappa_+,\kappa_-$ glue together to a morphism
\Beq
 \kappa: \L(H_r)\setminus\{0\}\to\mathbb R_+\times\mathbb S
\Eeq
and one shows by a brute force calculation
that $\iota\kappa=1_{\L(H_r)\setminus\{0\}}$,\ \
$\kappa\iota=1_{\mathbb R_+\times\mathbb S}$.
Hence \eqref{H=RxS} is indeed an isomorphism, and it follows that
$\epsilon$ is a regular closed embedding which makes $\mathbb S$ a
sub-supermanifold of $\L(H)$.

We claim that the composite
\Beq
 \omega: \L(H_r)\setminus0 \too\kappa
 \mathbb R_+\times\mathbb S \too{\pr_2} \mathbb S \too\epsilon \L(H_r)
\Eeq
is the "normalization morphism"
\Beq
 \hat\omega = \hat\omega[x] = \frac 1{\br\|x\|} x.
\Eeq
Indeed,
\Beq
 \hat\omega = \omega^*(x) = \kappa^*{\pr_2}^*\epsilon(x)
 = \kappa^*(\pm h + \frac 2{1+\br\|y_\pm\|^2}(y_\pm \mp h))
 = \pm h + \frac 2{1+\br\|\kappa^*(y_\pm)\|^2}(\kappa^*(y_\pm) \mp h);
\Eeq
investing \eqref{|Kap*y|}, the result follows after some more steps.

In view of \eqref{PullBIs0}, it remains to prove: Given some smf
morphism $\phi:Z\to\L(H_r)$ with
\Beq
 \phi^*(\br\|x\|^2-1) =0,
\Eeq
$\phi$ factors through $\mathbb S$. Indeed, we have
 $1= \phi^*(\br<\cj x|x>) = \Bigl<\cj{\hat\phi}|\hat\phi\Bigr>
 = \bigl\|\hat\phi\bigr\|^2$,
i.e. $(\bigl\|\hat\phi\bigr\|-1)(\bigl\|\hat\phi\bigr\|+1)=0$;
by Cor. \ref{InvOfSfct}, the second factor is invertible, and hence
\Beqn NPhi=1
  \bigl\|\hat\phi\bigr\|=1.
\Eeq
In particular, $\Im(\tilde\phi)\seq\tilde{\mathbb S}\seq\L(H_r)\setminus0$.
Now we claim that $\phi$ coincides with the composite
\Beq
 Z \too\phi \L(H_r)\setminus0  \too\omega \L(H_r)\setminus0.
\Eeq
Indeed, using \eqref{NPhi=1},
\Beq
 \widehat{\omega\phi} = \frac 1{\phi^*(\br\|x\|)}\phi^*(x) =
 \frac 1{\br\|\phi^*(x)\|}\phi^*(x) = \phi^*(x) = \hat\phi,
\Eeq
and now $\phi=\omega\phi=\epsilon\pr_2\kappa\phi$ provides the
factorization wanted.
\end{proof}

\Brm
The whole story becomes much more transparent if one looks at the
$Z$-valued points in the sense of \ref{Ber2Rog}.
Given any smf $Z$, we get from Thm. \ref{CoordChar} an identification
of the set $\L(H_r)(Z)$ of $Z$-valued points of $\L(H_r)$ with
$\M^{H_r}(Z)$, and thus the structure of a (purely even) vector
space on $\L(H_r)(Z)$. Of course, thus structure varies functorially
with $Z$.
Moreover, \eqref{The<>Pair} yields a bilinear map
$\L(H_r)(Z)\times\L(H_r)(Z) = \M^{H_r}(Z)\times \M^{H_r}(Z)\to \M(Z)$, i.~e.
a scalar product on $\L(H_r)(Z)$ with values in the algebra
$\M(Z)$.

Now the $Z$-valued points of $\mathbb S$ identify with those
$Z$-valued points of $\L(H_r)(Z)$ which have squared length equal to one
under this scalar product:
\Beq
 \mathbb S(Z) = \{\xi\in \L(H_r)(Z):\ \ \br<\cj \xi|\xi>-1=0;\}.
\Eeq
Now the process of stereographic projection can be applied in the vector
space $\L(H_r)(Z)$ for each $Z$, using the constant morphisms
$\pm h: Z\to P \too{\pm h} \L(H_r)$ as projection centers and the subspace
$\L(E)(Z)$ as screen. One gets maps
\Beq
 c^Z_\pm:
 \{\xi\in\mathbb S(Z):\ \  \text{$(\pm\Re\br<\cj\xi,h>-1)$ is invertible}\}=
 (\mathbb S\setminus\{\pm h\})(Z) \to \L(E)(Z)
\Eeq
given by the same formula \eqref{ActC}.

These maps vary functorially with $Z$ again; on the other hand, we noted
already in \ref{Ber2Rog} that the functor \eqref{X2X(cdot)} is faithfully full,
that is, "morphisms between representable
functors are representable". Thus, the maps $c^Z_\pm$ must be induced by
suitable morphisms, and these are indeed just our charts $c_\pm$. The
other morphisms $\iota,\epsilon,\kappa,\omega$ can be interpreted
similarly.
\Erm


\vfill

{\sc Technische Universit\"at Berlin}

Fachbereich Mathematik, MA 7 -- 2

Stra\ss e des 17. Juni 136

10623 Berlin

FR Germany

\medskip

{\em E-Mail  address: } schmitt@math.tu-berlin.de

\end{document}